\documentclass[manuscript,screen]{acmart}

\usepackage{framed}
\usepackage{multirow}
\usepackage{graphicx}
\usepackage{booktabs}
\usepackage{footnote}
\makesavenoteenv{tabular}
\makesavenoteenv{table}
\usepackage{xcolor}
\usepackage{makecell}
\usepackage[np]{numprint}

\def\doc{\includegraphics[height=0.015\textheight]{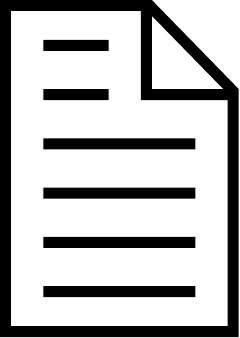}}
\def\ecom{\includegraphics[height=0.015\textheight]{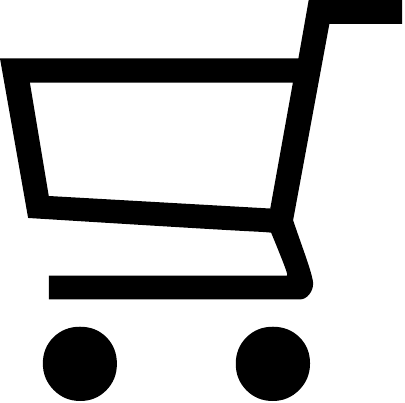}}
\def\energy{\includegraphics[height=0.015\textheight]{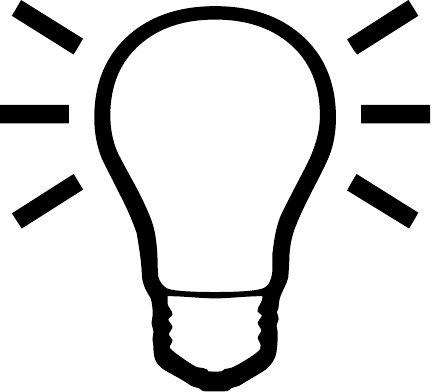}}
\def\education{\includegraphics[height=0.015\textheight]{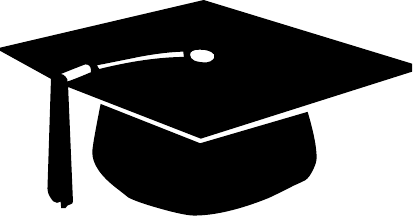}}
\def\health{\includegraphics[height=0.015\textheight]{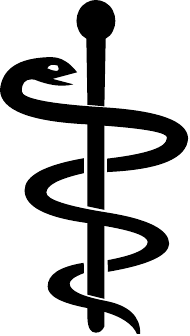}}
\def\movie{\includegraphics[height=0.015\textheight]{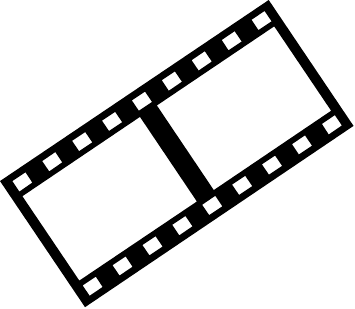}}
\def\music{\includegraphics[height=0.015\textheight]{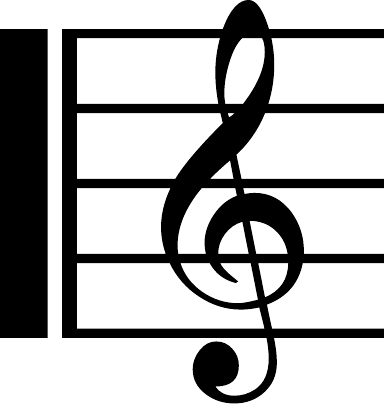}}
\def\dating{\includegraphics[height=0.015\textheight]{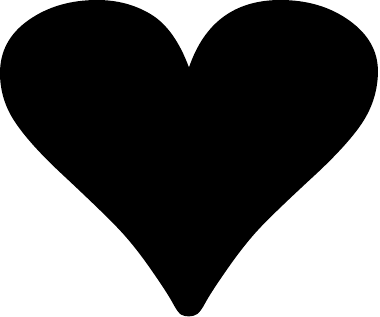}}
\def\poi{\includegraphics[height=0.015\textheight]{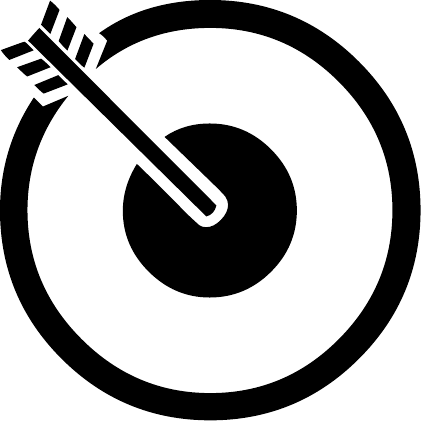}}
\def\social{\includegraphics[height=0.015\textheight]{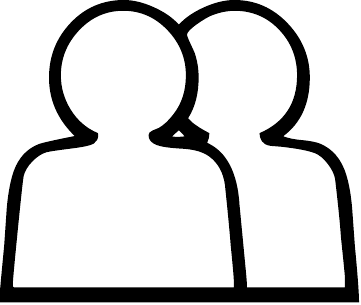}}

\acmJournal{TORS}

\begin{document}

\title[Whom do Explanations Serve?]{Whom do Explanations Serve? A Systematic Literature Survey of User Characteristics in Explainable Recommender Systems Evaluation}

\author{Kathrin Wardatzky}
\email{wardatzky@ifi.uzh.ch}
\orcid{0000-0002-7043-7326}
\affiliation{%
  \institution{University of Zurich}
  \city{Zurich}
  \country{Switzerland}
}

\author{Oana Inel}
\email{inel@ifi.uzh.ch}
\orcid{0000-0003-4691-6586}
\affiliation{%
  \institution{University of Zurich}
  \city{Zurich}
  \country{Switzerland}
}

\author{Luca Rossetto}
\email{luca.rossetto@dcu.ie}
\authornote{This work was done while the author was at the University of Zurich.}
\orcid{0000-0002-5389-9465}
\affiliation{%
  \institution{Dublin City University}
  \city{Dublin}
  \country{Ireland}
}

\author{Abraham Bernstein}
\email{bernstein@ifi.uzh.ch}
\orcid{0000-0002-0128-4602}
\affiliation{%
  \institution{University of Zurich}
  \city{Zurich}
  \country{Switzerland}
}

\renewcommand{\shortauthors}{Wardatzky, et al.}

\begin{abstract}
  Adding explanations to recommender systems is said to have multiple benefits, such as increasing user trust or system transparency.
  Previous work from other application areas suggests that specific user characteristics impact the users' perception of the explanation.
  However, we rarely find this type of evaluation for recommender systems explanations.
  This paper addresses this gap by surveying 124 papers in which recommender systems explanations were evaluated in user studies. 
  We analyzed their participant descriptions and study results where the impact of user characteristics on the explanation effects was measured.
  Our findings suggest that the results from the surveyed studies predominantly cover specific users who do not necessarily represent the users of recommender systems in the evaluation domain.
  This may seriously hamper the generalizability of any insights we may gain from current studies on explanations in recommender systems. 
  We further find inconsistencies in the data reporting, which impacts the reproducibility of the reported results.
  Hence, we recommend actions to move toward a more inclusive and reproducible evaluation.
\end{abstract}




\maketitle

\section{Introduction}
\label{sec:intro}

Recommender systems have become part of many people's everyday online interactions. 
Whether it is suggestions for what movie to watch, what items to buy, which news to read, or which restaurant to visit, these systems generally reach and serve a broad audience in their decision-making process.
Adding explanations to recommender systems is said to serve a multitude of benefits. First, by providing the reasoning behind specific recommendations, explanations can mitigate issues dealing with users' autonomy that can occur when recommendations nudge or persuade users in a particular direction that is not beneficial for them~\cite{milano2020recommender}. Carefully designed explanations can help the user to identify when the system is ``wrong'' and provide options to correct it (see the scrutability goal of \citet{tintarev_survey_2007}). Second, they can help users understand why they receive a specific recommendation, increase their trust, or allow them to make better decisions~\cite{tintarev_survey_2007, vultureanualbisi_survey_2022}.
The quality of recommender systems explanations is often assessed based on whether it has such effects on a user.
Measuring the explanation quality with offline metrics is still an open research challenge~\cite{zhang2020explainable}. Text-based explanations have been evaluated with metrics from the natural language processing domain. \citet{Ariza_2024_comparative}, for example, use BLEU-n, ROUGE-n, and BERT-S along with content repetition, explanation length, and sentence uniqueness measures to measure and compare the text quality of the generated explanations. 
In addition to BLEU and ROUGE, \citet{kokubo2022_explainable} and \citet{Yi2023_adarex} evaluate the diversity of the generated sentences in the explanation with the Distinct score. \citet{Yi2023_adarex} additionally use METEOR, a score based on unigram matching between the generated text and a ground truth.
These metrics are, however, limited to text-based explanations.
Other metrics include ranking- and retrieval-based metrics, such as Recall and NDCG used in \citet{Yi_2024_sequential}. 
Except for the diversity-based metrics, these metric types have the downside that they require a ground truth to compare the generated explanations, which is often approximated through online reviews.
The results from using online reviews as a ground truth need to be carefully discussed and interpreted as they can be biased~\cite{Hu_2017_review_bias} and might exclude certain sociodemographics and personality types~\cite{manner2017posts}.
\citet{Liao2020_questioning} further point out that the usefulness of explanations for users is dependent on their motivation for using the system and their downstream actions, which cannot be captured by current offline evaluation metrics.

Previous work found that user characteristics, such as personality, expertise, or cognitive abilities, lead to differences in the effect of explanations on a user.
Several works have shown that the Big Five personality trait can impact the persuasiveness of the recommender systems explanations~\cite{alslaity2020effect, sofia2016investigating, fatahi2023investigating}.
The characteristics can also impact how visual interfaces are perceived and interacted with~\cite{tintarev_effects_2016, conati_evaluating_2014}, or what type of transparency mechanisms for algorithmic decision-making are preferred~\cite{Springer_ProgressiveDisclosure}.
Users' cognitive abilities, such as perceptual speed, disembodiment, visual and verbal working memory, and personal characteristics, such as personality traits, impact, for example, the effectiveness of visualizations~\cite{conati2008exploring, sheidin2020effect}.
Furthermore, interfaces tailored to users' cultural backgrounds improve their satisfaction and efficiency with navigating it~\cite{Reinecke_culturally_adapted}.

The recommender systems community also has a long-standing history of investigating user characteristics to improve the quality of recommendations~\cite{hong2019cross} and the overall user experience (UX)~\cite{Tkalcic2015}.
However, \emph{only a few studies have been conducted in the explainable recommender systems community to investigate how different users perceive explanations}~\cite{naiseh_personalizing_xai_recsys}, which is why this issue is still widely unknown. This subject matter is, nevertheless, of particular importance, as recent research~\cite{langer2021we} suggests that the content, design, and goal of the explanations for typical decision-support systems should depend on all stakeholders' needs and interests.

One popular form of recruiting participants for human-subject evaluation is via crowdsourcing platforms such as Amazon Mechanical Turk (MTurk)\footnote{\url{https://www.mturk.com/}} or Prolific,\footnote{\url{https://www.prolific.co/}} who offer access to a global workforce. Research, however, has shown that their participant pool is not as diverse as advertised~\cite{difallah2018demographics}, crowd workers on MTurk being, on average, younger and with a higher income than the average population of their country. 
Further issues with crowdsourcing platforms were found regarding data quality and internal validity, including when measuring participants' personality~\cite{douglas2023data, chmielewski2020mturk}. While several user characteristics have been identified as relevant in typical user studies, such as age, gender, geographic location, ethnicity, language proficiency, education level, expertise, political, religious, and sexual orientation, physical and mental health, and disability~\cite{dong2012war,barbosa2019rehumanized,kapania2023hunt}, only a few are actually considered when selecting study participants, namely geographic location and language proficiency~\cite{kapania2023hunt}. Furthermore, research has shown that participant samples in human-subject experiments often come from Western, Educated, Industrialized, Rich, and Democratic (WEIRD) societies~\cite{henrich_heine_norenzayan_2010, Sturm_weirdHCI, Linxen_2021_weirdCHI}, which means that \emph{the results of these experiments only represent a subset of the overall population}.

In this work, \emph{we systematically analyze what we know about the participants of studies involving recommender system explanations regarding user characteristics such as demographics, prior knowledge, and personality traits}.
We look at publications where the effects of explanations were measured with respect to different user characteristics.
Often, these effects equal the goals or benefits of adding explanations to recommender systems as identified in \citet{tintarev_survey_2007, Tintarev2015}.
One challenge that we face is that, to this date, there are no commonly accepted standards or guidelines for evaluating explanations.
This led to differences in the evaluation approaches and made conducting a meta-analysis impossible.
Therefore, \emph{our research regarding the explanation effects focuses on exploring \textbf{which (user) characteristics have been analyzed for which explanation effect}}.

To achieve these goals, we systematically surveyed 124 peer-reviewed papers published between 2017 and 2022 in which explanations were evaluated in user studies. 
We analyzed them to answer the following research questions:

\begin{description}
    \item[RQ 1:] Who was recruited to evaluate recommender systems explanations?
    \item[RQ 2:] Which user characteristics impact the effect of explanations in recommender systems?
\end{description}

In the first research question, we look at the description of the participants in the user studies.
We analyze the participants' demographic information, personality traits, and prior experience to provide more context to the evaluated explanation effects.
Subsequently, we discuss what user groups have not been included when evaluating recommender systems explanations based on the results of RQ 1.
By creating awareness about these gaps, we hope to generate opportunities to specifically guide research endeavors in a direction that has improved coverage of the entire range of recommender system users.
For the second research question, we look at the findings regarding the explanations' effects on users and investigate how these might differ for specific user characteristics.
We look at and discuss the papers to show possible indications for the effects that have been evaluated and highlight gaps where more research is needed.\footnote{We provide an interactive version of this analysis here: \url{https://kathriwa.github.io/interactive-survey-visualization/\#/}.}

We conclude with recommendations regarding participant recruitment, reporting of participant data, and evaluation of explanation effects.
Hence, in addition to highlighting the interaction between user characteristics and explanations, this work also aims to raise awareness about reproducibility, result reuse, and inclusion issues related to evaluating recommender systems explanations and spark a discussion about addressing these issues in the research community.

\section{Related Work}
\label{sec:related-work}

In the literature, several surveys mention or review aspects of the participants in human-subject evaluation of recommender systems explanations or the effects these explanations have on the user. The literature survey by \citet{naiseh_personalizing_xai_recsys} is probably closest to this work.
They surveyed 48 publications on user needs and implementations of personalized explanations in recommender systems.
The authors categorize the findings into users' motivation to interact with the system, their goals to examine the explanations, cognitive load, decision cost, and regulation compliance.
They conclude that more user-based research is needed to learn more about how their perception and preferences relate to aspects such as personality, domain knowledge, and user goals.

In terms of general overviews about explainable recommender systems, \citet{zhang2020explainable} point out the need for user behavior analysis and user perspectives as recommender systems are \emph{``inherently human-computer interaction systems''.}
\citet{nunes2017systematic} looked at the number of participants for different study designs in the evaluation of explainable decision support and recommender systems.
They also investigated the dependent variables evaluated in their corpus but did not connect these findings with the participants. Several surveys focus on the design and evaluation of recommender systems explanations.
For example, \citet{Tintarev2015} focus their investigations on the measured effects or goals that explanations are evaluated on but do not provide insights with regard to the participants conducting the evaluations.
\citet{mohseni2021multidisciplinary} follow a similar route and analyze design goals for AI explanations.
They relate these goals to the targeted user type, which they classify based on their experience into AI novices, data experts, and AI experts.

Outside of the recommender systems community, \citet{chromik2020taxonomy} propose a general taxonomy for human-subjects evaluation for explainable AI.
The participant dimension of their taxonomy includes aspects concerning the study type and design, such as the number of participants, their incentivization to participate, how they were recruited, and the participants' foresight (i.e., is the study assuming that all participants have the same knowledge about the context or can they draw from external facts, such as prior experiences).
The taxonomy also includes the participants' AI and domain expertise but no other user characteristics.
In contrast, \citet{nauta_anecdotal_2023} do not focus their survey on the user-based evaluation. 
Still, they point out that in their corpus spanning the years 2014-2020, only one in five papers evaluate the explanations with users.
Their analysis focuses on 12 properties for good explanations and how these can be evaluated quantitatively but do not include user-based evaluations.

Overall, the participants in human-subject evaluation of AI explanations and the effects that explanations can have on the users have been on the radar of some surveys, but these dimensions have not been connected yet.

Aside from survey papers, the impact of the Big Five personality traits on the persuasiveness of recommendations using explanations designed according to Cialdini's six persuasive principles (reciprocity, scarcity, authority, social proof, liking, and commitment)~\cite{cialdini2009influence} has been evaluated by \citet{alslaity2020effect}, \citet{sofia2016investigating}, and \citet{fatahi2023investigating}.
\citet{alslaity2020effect} found differences between the two evaluated domains---e-commerce and movie recommendations---within the same personality trait group for the majority of the persuasion profiles.
They also found statistically significant interactions between personality traits and domain for three persuasive principles (reciprocity, liking, and scarcity).
\citet{sofia2016investigating} designed justifications following Cialdini's persuasive principles for a music recommender system to promote new artists' songs.
They not only found differences in the reception of the justifications between participants with different personalities but also that the participants were not very good at assessing which justification type would be the most persuasive for them.
\citet{fatahi2023investigating} aimed to use the explanations to persuade users of movies that they were initially unmotivated to watch.
They show that explanations containing influence strategies that are tailored to the respective user personality can successfully persuade them to interact with items that were previously not of interest and plan to extend the analysis to other personality traits, such as the need for cognition.

These findings underline the motivation for this paper to systematically analyze which users were recruited and investigate what is known about the potential impacts of user characteristics on the measured explanation effects. 

\section{Methodology}
\label{sec:method}

This section describes the paper's methodology by explaining the data collection, selection, and annotation process.

\subsection{Data Collection and Selection}
\label{sec:data-collection}
We conducted a systematic literature survey of peer-reviewed papers published between 2017 and 2022 to answer the research questions outlined in Section \ref{sec:intro}.
We opted to start the collection with the year 2017 as this year seems to have been the starting point of a steadily increasing number of publications in the explainability field \cite{vultureanualbisi_survey_2022, barredo_arrieta_explainable_2020}.
To collect the papers, we queried six library databases (see Table \ref{tab:libraries}) with the following search term:
\begin{framed}
\noindent \texttt{(explain$\ast$ OR explanation$\ast$ OR interpretab$\ast$ OR intelligib$\ast$ OR justification OR transparen$\ast$) AND \\(recommender OR recommendation OR personalization OR personalized)}
\end{framed}
We constructed the search term to contain words frequently used interchangeably to refer to both explainable AI and recommender systems or recommendations.
The first part of the search term includes terms related to the explanation part. 
Given that there are currently no commonly agreed-upon definitions of AI explanations \cite{meske_explainable_2022, guidotti_survey_2019}, we included additional terms aiming to capture potentially relevant articles using a different terminology along with variants of \emph{explainability} and \emph{interpretability}.
Some articles distinguish between \emph{justifications} and \emph{explanations} while others use \emph{explanation} for \emph{justifications}. 
Therefore, we opted to include \emph{justifications} in the search. 
The variations of \emph{transparency} aim to include those articles that opt for a different wording.

The six library databases were selected to cover major computer science venues and interdisciplinary outlets.

\begin{table}
\centering
    \caption{Library databases and the number of search results.}
    \label{tab:libraries}
    \Description{Overview of the library databases we queried with our search term and the number of results we received. At the bottom of the table, we sum the results and state how many papers were selected for our corpus.}
    \begin{tabular}{lr}
        \toprule
        \textbf{Database} & \textbf{\# Results} \\
        \midrule 
        ACM Digital Library\footnote{\url{https://dl.acm.org/}}  & \np{905} \\
        IEEEXplore\footnote{\url{https://ieeexplore.ieee.org}} & \np{966} \\
        Taylor \& Francis\footnote{\url{https://www.tandfonline.com/}} & \np{1808}  \\
        Web of Science\footnote{\url{https://www.webofscience.com}} & \np{13808}  \\
        Wiley Online\footnote{\url{https://onlinelibrary.wiley.com/}} & \np{2467}  \\
        SemanticScholar\footnote{\url{https://www.semanticscholar.org/}} & \np{110000}   \\
        \midrule
        Total Retrieved Results & \np{129954}   \\
        \textbf{Final Corpus Size} & \textbf{124} \\
        \bottomrule
    \end{tabular}
\end{table}

This initial search resulted in a total of \np{129954} returned articles.
In the first round, we filtered out all papers that, based on title and abstract, were clearly not related to explainable AI or recommender systems.
Second, we further filtered using the exclusion and inclusion criteria (EC and IC) summarized in Table~\ref{tab:selection}.
Hereby, we excluded all publications that were not written in English and where we could not access the full text through the licenses of our institution (EC-1 and EC-2).
Furthermore, we excluded papers with results published in another paper matching our inclusion and exclusion criteria (EC-3). We kept the extended paper in our considered set of papers.
Duplicated papers that appeared in multiple databases were only considered once (EC-4).

The remaining papers had to fulfill all following inclusion criteria to be selected.
We included papers that propose an explainable or interpretable recommender system or an explanation generation method (IC-1), and the explanations are evaluated in a recommender systems context (IC-2).
To ensure that the explanations are a focus of the papers, we only included publications providing examples or detailed descriptions of the explanations (IC-3).
The final two inclusion criteria refer to the evaluation method.
Our corpus includes papers in which the explanations were evaluated with a user study (IC-4), where the results were tested for statistical significance (IC-5).

\begin{table}
    \caption{Exclusion and inclusion criteria for the paper selection process.}
    \Description{List of three exclusion and five inclusion criteria that were applied in the paper selection process.}
    \label{tab:selection}
    \resizebox{1\textwidth}{!}{
    \begin{tabular}{ll}
        \toprule
        \multicolumn{2}{l}{Exclusion Criteria (EC)} \\
        \midrule
        EC-1 & The paper is not written in English \\
        EC-2 & We have no access to the full paper \\
        EC-3 & The content of the paper was published in an extended paper that matches the inclusion criteria \\
        EC-4 & The paper was already retrieved from another database \\
        \midrule
        \multicolumn{2}{l}{Inclusion Criteria (IC)} \\
        \midrule 
        IC-1 & The paper proposes an explainable or interpretable recommender system or an explanation generation method\\
        IC-2 & The paper presents explanations that are evaluated in a recommendation scenario \\
        IC-3 & The paper provides an example or detailed description of the explanation \\
        IC-4 & The explanations are evaluated on users \\
        IC-5 & The results were tested for statistical significance \\
        \bottomrule
    \end{tabular}
    }
\end{table}

The selection process narrowed the corpus to 124 papers containing a total of 158 user studies (some papers reported on multiple studies), from which we extracted the information reported on the user study participants and the results.\footnote{The full list of papers and categorization can be viewed here: \url{https://doi.org/10.5281/zenodo.14771123}} In Section~\ref{sec:data_annotation}, we elaborate on the annotation process of the papers in our corpus.

\subsection{Data Annotation}
\label{sec:data_annotation}
We extracted general information on the application domain, the evaluation methodology, and the explanation that was evaluated.

To analyze the first research question, we focused on the participant descriptions from the 158 user studies in the papers of our corpus.
We extracted all user characteristics mentioned and, if provided, the participants' distribution.
The characteristics were then sorted into demographic, personality, and experience categories as suggested by \citet{EGAN1988543}. 
Table~\ref{tab:recorded-characteristics} provides an overview of the extracted characteristics.

To answer the second research question, we annotated the findings in the results sections of each paper. More precisely, we extracted the dependent and independent variables of each finding in which a user characteristic was part of a variable along with the outcome of the evaluation (i.e., if an effect was found or not).
The dependent variables, or measured explanation effects, were then categorized in the next step.
We noticed that the effects did not consistently follow the same definitions and were frequently named differently.
Therefore, we referred to the evaluation task or question used for the evaluation, wherever possible (i.e., when they were made available by the authors of the papers).
We categorized the explanation effects by the seven explanation goals defined by \citet{tintarev_survey_2007}: effectiveness, efficiency, transparency, persuasiveness, trust, satisfaction, and scrutability.
By looking at the user task design, we noticed that the papers stating to evaluate the scrutability of the system were doing so by assessing the perceived system control.
We, therefore, combined the goals of scrutability and satisfaction, satisfaction being defined as \emph{``increase the ease of usability or enjoyment''}~\cite{tintarev_survey_2007}, into a usability/UX category.
Aside from these goals, we also observed frequent evaluations of the users' perceived explanation quality. 
Given that user characteristics might impact the user's perception, we added it as an additional factor to the effects we investigated.

As this analysis is a first step to gaining insights into how user characteristics interact with explanation effects, we omitted results for effects specific to the recommendation problem (e.g., group recommendations) or the application domain (e.g., change in driving behavior in autonomous driving).

\section{Results}
\label{sec:results}
In this section, we present the results of the data analysis to answer our two research questions.
We analyze the results and participants' information from the 158 user studies conducted in the 124 papers. 
First, we look at the information we could extract regarding study participants. 
Then, we analyze the user study findings in which the effects of explanations were measured by disaggregating different user characteristics.

\subsection{Participants}
\label{sec:participants}
The participants in the user studies presented in our corpus were primarily recruited on crowdsourcing platforms such as MTurk or Prolific and at universities.

Table \ref{tab:recorded-characteristics} provides an overview of the extracted characteristics, the number of papers reporting this information about their participants, and the number of papers evaluating whether the given user characteristic impacts the explanation effect.

\begin{table}
\caption{Overview of the recorded user characteristics reported in the papers and the papers in which the characteristics were analyzed.}
\label{tab:recorded-characteristics}
\Description{The first two columns of the table list the user characteristics we identified in our corpus sorted by their category (demographic, personality, experience).
The third column lists the references to all papers that recorded information, along with the number of studies stated in the fourth column.
Columns five and six repeat this for the papers that analyzed the characteristics in their evaluation.}
\resizebox{\columnwidth}{!}{%
\begin{tabular}{@{}llp{8cm}lp{3cm}l@{}}
\toprule
\makecell{Type of \\ Characteristic} & Characteristic & Papers Recording Characteristic & \makecell{\# Studies Recording \\ Characteristic} & \makecell{Papers Analyzing \\ Characteristic} & \ \makecell{\# Studies Analyzing \\ Characteristic} \\ \midrule
\multirow{13}{*}{Demographic} & Age & \cite{paper2,paper1,paper4,paper5,paper6,paper7,paper10,paper12,paper13,paper15,paper17,paper18,paper19,paper20,paper21,paper22,paper23,paper24,paper25,paper26,paper27,paper28,paper29,paper30,paper33,paper34,paper35,paper36,paper39,paper40,paper42,paper43,paper47,paper52,paper53,paper54,paper55,paper57,paper58,paper59,paper60,paper61,paper62,paper63,paper65,paper67,paper68,paper69,paper71,paper74,paper75,paper76,paper77,paper78,paper81,paper82,paper83,paper84,paper86,paper87,paper88,paper89,paper90,paper91,paper92,paper93,paper95,paper97,paper98,paper99,paper100,paper102,paper103,paper105,paper106,paper107,paper108,paper109,paper111,paper112,paper114,paper118,paper119,paper120,paper121,paper122,paper123,paper125} & 114 & \cite{paper21,paper58,paper111,paper125} & 4 \\
 & Gender & \cite{paper2,paper1,paper4,paper5,paper6,paper10,paper12,paper13,paper15,paper17,paper18,paper19,paper20,paper21,paper22,paper23,paper24,paper25,paper26,paper27,paper28,paper29,paper30,paper33,paper35,paper36,paper39,paper40,paper42,paper43,paper46,paper47,paper52,paper53,paper54,paper55,paper57,paper58,paper59,paper60,paper65,paper68,paper69,paper71,paper74,paper75,paper76,paper77,paper78,paper81,paper82,paper83,paper84,paper87,paper88,paper89,paper90,paper91,paper92,paper93,paper95,paper97,paper98,paper99,paper100,paper102,paper103,paper105,paper107,paper108,paper109,paper111,paper112,paper114,paper115,paper118,paper119,paper120,paper121,paper122,paper125} & 105 & \cite{paper21,paper29,paper58,paper60,paper97,paper99,paper111,paper125} & 7 \\
 & Location & \cite{paper2,paper4,paper6,paper7,paper9,paper13,paper14,paper18,paper19,paper21,paper22,paper23,paper34,paper36,paper38,paper40,paper42,paper44,paper45,paper47,paper50,paper52,paper53,paper54,paper56,paper61,paper62,paper63,paper64,paper68,paper69,paper71,paper75,paper77,paper78,paper79,paper83,paper86,paper87,paper88,paper92,paper94,paper96,paper97,paper98,paper99,paper100,paper101,paper102,paper103,paper104,paper106,paper109,paper110,paper111,paper112,paper114,paper118,paper121,paper125} & 74 & \cite{paper103} & 1\\
 & Education & \cite{paper2,paper1,paper4,paper4,paper14,paper18,paper19,paper20,paper21,paper22,paper23,paper24,paper26,paper29,paper39,paper40,paper42,paper44,paper45,paper46,paper47,paper54,paper56,paper57,paper59,paper64,paper71,paper77,paper79,paper81,paper89,paper90,paper94,paper95,paper96,paper102,paper107,paper110,paper111,paper112,paper114,paper115,paper118,paper119,paper120,paper122,paper123} & 51 & \cite{paper29,paper41,paper99,paper111} & 4 \\
 & Other & \cite{paper1, paper2, paper11,paper17, paper26, paper33, paper36, paper41, paper60, paper82,paper93, paper115, paper116,paper119, paper121, paper125} & 22 & - & 0 \\
 \midrule
\multirow{14}{*}{Personality} & Need for Cognition & \cite{paper10,paper12,paper23,paper42,paper57,paper74,paper84,paper115} & 11 & \cite{paper10,paper12,paper18,paper23,paper57,paper74,paper84} & 8 \\
 & Openness & \cite{paper10,paper21,paper22,paper27,paper43,paper65,paper84,paper87} & 10 & \cite{paper22,paper87} & 3 \\
 & Conscientiousness & \cite{paper21,paper22,paper27,paper43,paper65,paper84,paper87} & 7 & \cite{paper21,paper22, paper87} & 4 \\
 & Extraversion & \cite{paper21,paper22,paper27,paper43,paper65,paper84,paper87} & 8 & \cite{paper21,paper22,paper65} & 3 \\
 & Agreeableness & \cite{paper21,paper22,paper27,paper43,paper65,paper84,paper87} & 9 & \cite{paper22,paper27,paper65} & 3 \\
 & Neuroticism & \cite{paper21,paper22,paper27,paper43,paper65,paper84,paper87,paper115} & 8 & \cite{paper21,paper22,paper27,paper65,paper87} & 5  \\
 & Innovativeness & \cite{paper18,paper23} & 2 & \cite{paper18,paper23} & 2 \\
 & Rationality & \cite{paper68,paper107} & 2 & \cite{paper68,paper107} & 2 \\
 & Social Awareness & \cite{paper36,paper68,paper69,paper75} & 5 & \cite{paper36,paper68,paper69,paper75} & 5 \\
 & Propensity to trust others & \cite{paper21,paper23,paper26,paper29,paper58,paper69,paper122} & 7 & \cite{paper21,paper23} & 2 \\
 & Trust in Technology & \cite{paper108} & 2 & \cite{paper108} & 1 \\
 & Valence & \cite{paper88} & 1 & \cite{paper88} & 1  \\
 & Decision-Making Type & \cite{paper68,paper69,paper75,paper86,paper97} & 7 & \cite{paper69,paper75,paper86,paper97} & 7 \\
 \midrule
\multirow{5}{*}{Experience} & Domain Experience & \cite{paper4,paper10,paper13,paper18,paper19,paper20,paper23,paper33,paper34,paper35,paper59,paper65,paper74,paper81,paper82,paper84,paper85,paper87,paper89,paper95,paper98,paper102,paper108,paper115,paper125,paper2,paper11,paper28,paper39,paper47,paper93,paper106,paper109,paper118,paper120,paper122,paper7,paper58,paper92} & 39 & \cite{paper23,paper74,paper84,paper87,paper108,paper109,paper125} & 7 \\
 & Technical Expertise & \cite{paper13,paper23,paper24,paper26,paper29,paper34,paper39,paper43,paper46,paper74, paper87,paper89,paper95,paper103} & 14 & \cite{paper18,paper23,paper34, paper87} & 4 \\
 & Visualization Literacy & \cite{paper18,paper23,paper69,paper74,paper75,paper87,paper122} & 8 & \cite{paper18,paper23,paper69,paper74,paper75,paper87} & 6 \\ \bottomrule 
\end{tabular}%
}
\end{table}

\subsubsection{Demographics}
\label{sec:result-dems}

\begin{figure}
    \centering
    \includegraphics[width=0.95\textwidth]{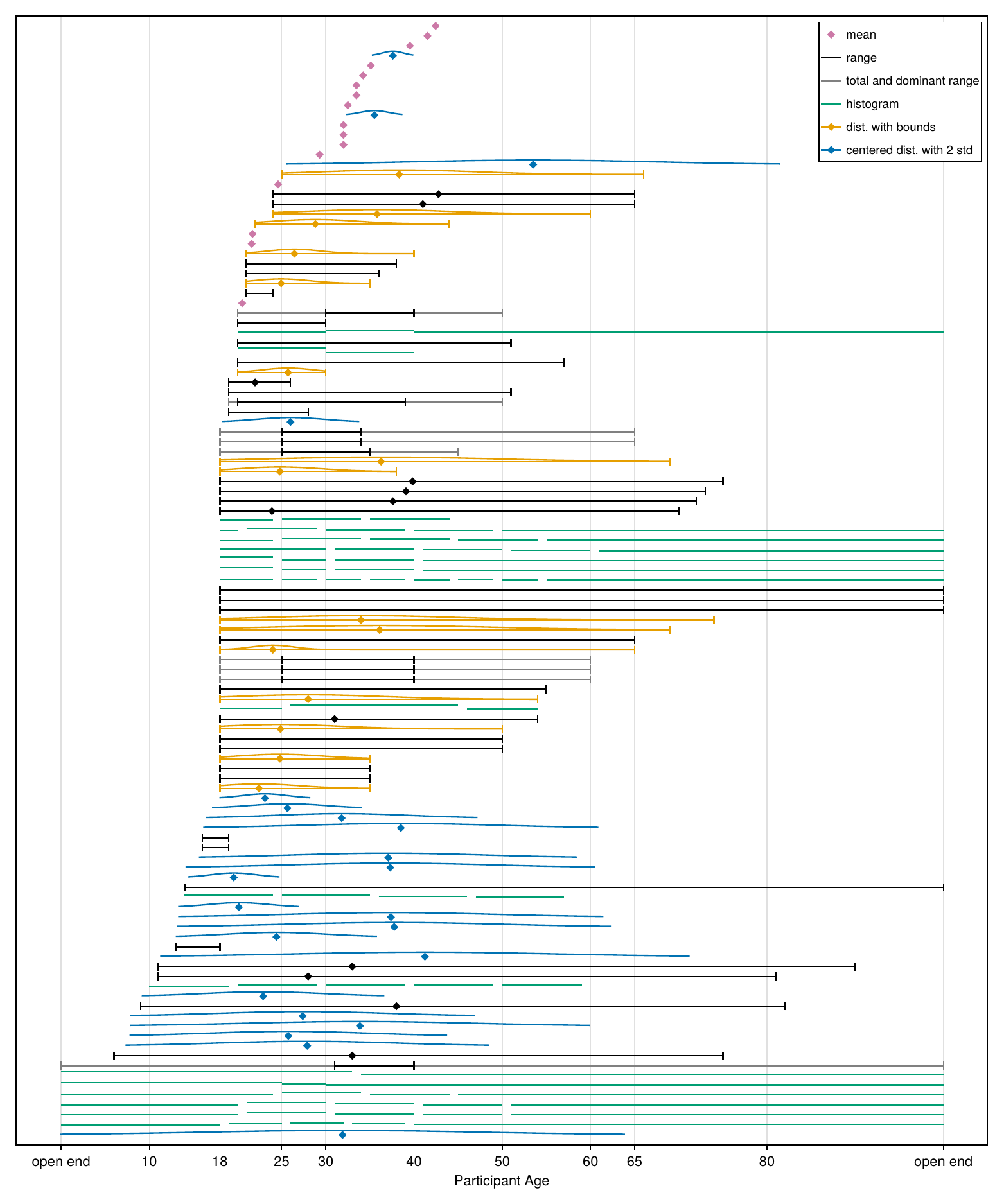}
    \caption{Different ways of reporting the participants' age. Each visualization corresponds to how the information was reported in the papers, where one row represents one study sorted by lower-bound age. Black bars denote age ranges, and combined black and gray bars denote total and `dominant' ranges. Blue curves denote centered distributions with two standard deviations around a mean. Yellow curves denote distributions with explicit lower- and upper bounds. Green lines represent explicit age histogram buckets, and diamonds $\blacklozenge$ represent mean values.}
    \Description{Plot where each line represents the information about the participants' ages for each study. For most studies, the average age was reported predominantly between 20 and 40.}
    \label{fig:participant-age}
\end{figure}

As can be seen in Table \ref{tab:recorded-characteristics}, it is common practice to provide demographic information about the study participants.
Only 18 of the 159 studies in our corpus did not disclose any demographic information about their participants.
User demographics reported for most of the studies are the age (114 studies) and gender (105) of the participants.
Other demographics that are less frequently mentioned are the country of residence (74), educational background (51), and ethnicity (3).

\paragraph{Age}
Participants' age is reported in multiple ways in the reviewed studies.
It was either stated as average (39 studies), distribution of age brackets (16), the entire age range of all participants (13), a combination of average age and the entire range (27), the majority age range (5), a combination of the entire age range and the majority age range (9), the lower age threshold (4), or as the percentage of participants younger than an age threshold (1).
Figure~\ref{fig:participant-age} summarizes the reported age values and depicts the diversity in reporting.
Depending on the information reported in a paper, the figure uses a different visualization for each presented age distribution.
The least detailed information is only a mean-average age value, visualized using a diamond symbol $\blacklozenge$.
This symbol can be combined with other information, such as a lower- and upper-bound, shown as a black interval.
When two intervals are reported in a paper, referring to a combination of the entire range and a majority range, the former is shown as a grey interval.
For papers that report a mean age and a standard deviation, Figure~\ref{fig:participant-age} shows a normal distribution centered around the mean, extending to $\pm2$ standard deviations, using a blue color.
When, in addition, explicit lower and upper bounds are reported, the distribution is capped at these values and shown in yellow.
The line segments shown in green represent explicit age brackets.
Whenever no explicit lower- or upper bound is given for a range or a collection of age brackets, the figure caps the lower-most at 0 and the upper-most at 100.
Since these values are not based on reported information, they are labeled `open end' in the figure.
The visualizations are sorted by (actual or inferred) lower bounds of the stated ranges.

Due to these differences, it is impossible to determine the average or standard age of the participants, but we can see certain tendencies.
Most user studies reported an average age between 20 and 40 years.
Underage participants are rarely covered, while studies including participants above 60 are more frequent; exact participant numbers are seldom explicitly specified for the upper age range.
Instead, we often find all participants above the age of a certain threshold grouped together, and it is unclear where the overall age range ends.
\citet{paper125} reported 13 of 123 participants above the age of 50, and \citet{paper58} had 13 of 310 participants above the age of 55.
In both cases, we cannot say if all the participants in this age bracket were 51 or 56, respectively, or if they were significantly older.
These open brackets of the oldest participants vary in their starting age as well. While most start in the early 50s, \citet{paper97}, for example, start at the age of 41, and the oldest recruited participants in \citet{paper89} are \emph{``above 30''}.

\paragraph{Gender}
We observed that most studies (93) reported the participants' gender in a binary way.
We either see a statement of the number or percentage of participants identifying as one gender (45) or the number or percentage of both male and female participants (48).
Out of these, seven studies reported participants who did not disclose their gender identity.
Eleven studies reported options for non-binary, diverse, self-described, non-listed, or other gender identities.
These participants are in the minority, with most studies reporting under 3\% of the overall participants not identifying as male or female.
Regarding the balance between male and female participants, we found that 55 of the 104 studies that reported the gender had a ratio within 10\% difference in participant numbers.
We found 49 studies that reported a difference between male and female participants higher than 10\% of the number of participants identifying with a binary gender; 30 studies had a male majority, while 19 had a female majority.
One study~\cite{paper103} reported their gender distribution as ``comparable distribution across countries in terms of gender'' from which we were not able to determine the options they provided in their questionnaire.

\begin{figure}
    \centering
        \includegraphics[width=\textwidth]{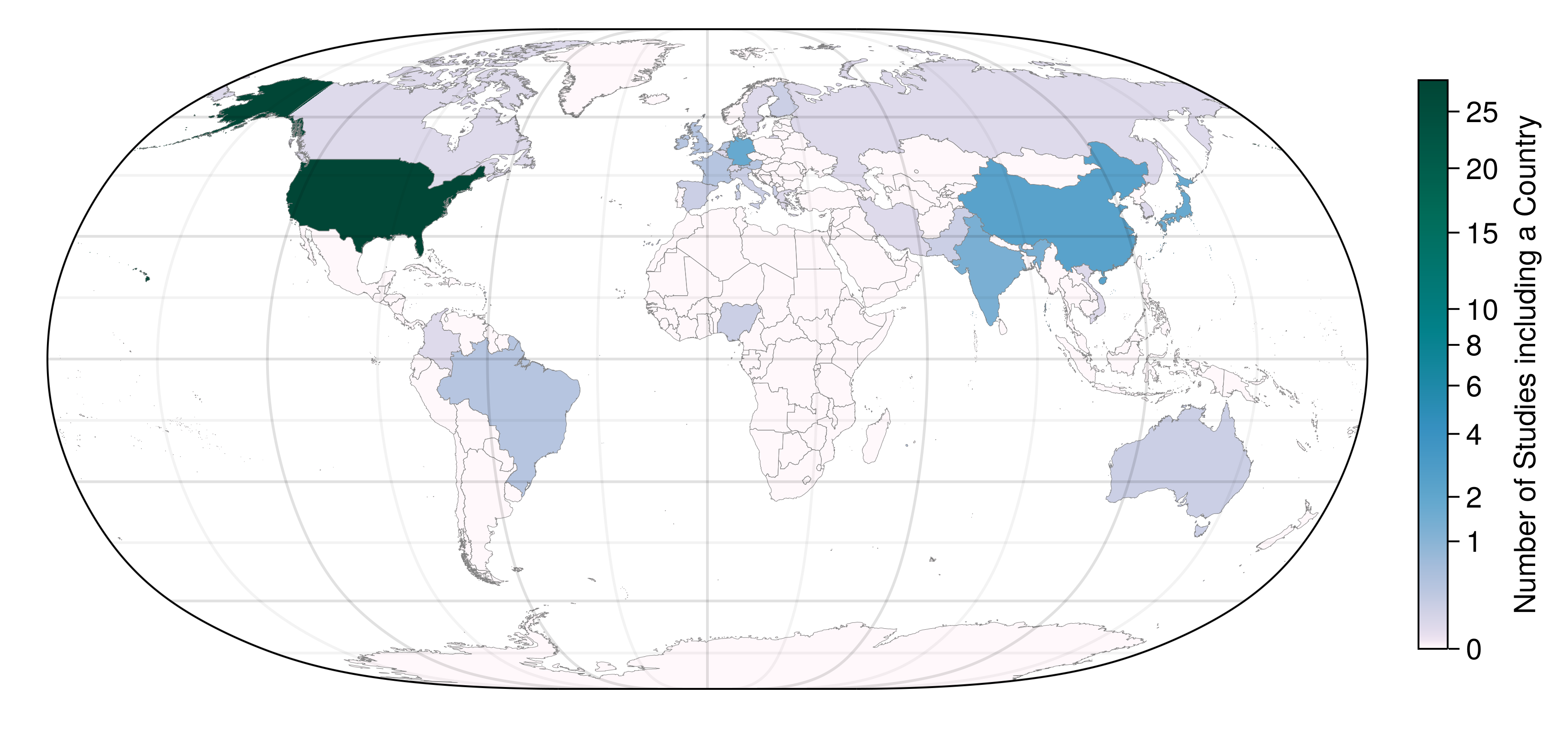}
        \caption{Number of studies involving participants from different countries}
        \label{fig:participantlocation}
        \Description{World map where the countries are colored depending on the number of studies that included participants from that country. The scale ranges from 0 to 28 studies per country.}
\end{figure}

\paragraph{Location}

Seventy-four studies reported the location of the participants.
These \np{8446} participants came from 28 different countries, covering 40\% of the total number of participants in our corpus.
Figure \ref{fig:participantlocation} depicts the number of studies in which a participant's country of residence was mentioned.
We notice that Africa, Central and South America, Eastern Europe, and South-East Asia are poorly represented.
The USA is the country where, by far, most studies have recruited participants.
Twenty-four studies recruited participants in the US, followed by China with six studies.

The participants for 24 studies were recruited from multiple countries.
The highest number of countries stated in a paper was 25 countries~\cite{paper52}.
This is the exception, though.
In studies with multi-national participants, the majority recruited participants from two to five countries.
However, not all papers reported the location of the participants on a country level but as aggregated. For example, 
\citet{paper22} described the location of their participants on a broader regional level (Oceania, South-East Asia, Northern-Western Europe), \citet{paper53} reported the number of participating countries. In contrast, \citet{paper62} used a country descriptor (e.g., English-speaking countries).

We further analyzed the distribution of participants across the countries.
Overall, we could extract the location for \np{7646} participants, meaning that 800 participants were listed among ``other'' or the exact distribution was not reported in the paper. In terms of numbers, the USA dominates with \np{3381} recruited participants. In the country with the second most participants, Japan, only \np{967} participants were recruited.
In fact, the sum of participants from the countries with second to ninth most participants is \np{3391}, only 10 participants more than the US.

\paragraph{Education}
The educational background of the participants is mainly associated with their highest degree.
Out of the 53 studies in which the participants' educational background was addressed, 28 did not disclose any distribution.
In most cases (22), the educational background is homogeneous as the researchers exclusively recruited students.
The remaining six studies provide a general overview of the educational background, a lower threshold which the majority passed (\emph{``almost all attained their high school diploma''}~\cite{paper63}), or a general description of the educational degrees \emph{``they also have different education majors (IT, education, finance, sociology, literature, mathematics, biology, etc.), degrees (high school, associate degree, Bachelor, Master, PhD)''~\cite{paper4}}).
All studies that reported information about the distribution predominantly recruited participants pursuing or having at least a Bachelor's degree.

\paragraph{Other Demographics}
We found the following demographic information that was only reported by a few studies: (1) the ethnic or cultural background of the participants, (2) their occupation or employment status, (3) the income level, (4) the participants' sexuality, (5) their marital status, (6) the household income, (7) the participants' body mass index, (8) their housing situation, and (9) whether they have a driver's license.

We found two papers that reported the distribution of the ethnic background of their participants~\cite{paper26, paper115}.
Both studies had predominantly Caucasian participants.
A third study~\cite{paper54} stated that their participants had a diverse cultural background without going into more detail.
Twelve studies recorded participants' occupations, but no information on the participants' distribution over the occupations was provided.
The remaining demographics were recorded when they were relevant to the application domain.
In online dating scenarios, studies reported the income level~\cite{paper121}, sexuality~\cite{paper60,paper1}, and marital status of the participants~\cite{paper1, paper60}.
The household income was also recorded in an e-commerce setting~\cite{paper2}. 
The body mass index~\cite{paper33} and employment status~\cite{paper11} were recorded in studies in food recommendation settings.
A study on electricity saving recommendations~\cite{paper36} recorded the housing situation of the participants, and a study collected information on the participants' driver's licenses on recommendations during autonomous driving~\cite{paper125}.

\subsubsection{Personality}
We found 36 studies in which information about the participants' personalities was reported across 27 publications.
Table~\ref{tab:recorded-characteristics} provides an overview of the recorded personality traits reported in our corpus.
These traits, outlined below, are often evaluated using standardized questionnaires.
The most frequent personality traits evaluated were the need for cognition, the so-called Big Five or OCEAN characteristics (openness, conscientiousness, extraversion, agreeableness, and neuroticism), and the participant's propensity to trust others.

\paragraph{Need for Cognition}
Need for Cognition (NFC) is the \emph{``Tendency for an individual to engage in and enjoy effortful cognitive activities''}~\cite{cacioppo1984nfc}.
Eleven studies (8 papers) stated that they measured the NFC of their participants.
We found that NFC is frequently evaluated by splitting the participants' scores by the median to form two groups with high and low NFC, respectively.
Hence, the reported distribution of participants is approximately equal.
\citet{paper84} was the only paper in which the average NFC score was reported instead of the distribution.

\paragraph{Big Five}
Nine of the eleven studies that reported Big Five or OCEAN personality traits evaluated all five traits.
The exceptions were \citet{paper10}, who focused on openness, and \citet{paper115}, who recorded the anxiety level of their participants, which can be classified as part of the neuroticism trait.

We found different approaches to report the participant distribution.
Two studies stated the mean and standard deviation of the respective scores~\cite{paper21, paper84}, both studies in \citet{paper22} provided the distribution of participants in low, medium, and high score-regions of each trait, and \citet{paper10} performed a median split and grouped all participants in low and high openness groups.
The remaining six studies did not provide details on the distribution.

\paragraph{Trust Propensity}
Trust propensity is defined as \emph{``Level of intensity of an individual's natural inclination to trust other parties in general''}~\cite{komiak2003trustpropensity}.
Eight studies (8 papers) looked into this personality trait. 
However, we could only extract the distribution across participants for one paper.
\citet{paper122} reported that the participants were approximately evenly distributed across high, medium, and low propensity scores.

\paragraph{Other Personality Traits}
We identified six other personality traits for which no participant distributions were reported.
Seven studies looked at the decision-making style of the participants.
We found two different dimensions concerning participants' decision-making strategy in our corpus.
Four studies from three papers classify their participants as rational or intuitive decision-makers.
The remaining three studies split their participants into people who make decisions to maximize the outcome (maximizers) and satisficers who settle for the first choice that is good enough.
We did not find further information on how the participants were distributed across the Social Awareness (4 studies), Personal Innovativeness (2 studies), Trust in Technology (1 study), and Valence (1 study) traits.

\subsubsection{Experience}
\label{subsubsec:experience}
We identified 53 studies in which the participants' prior experience was recorded.
The notions of experience that we observed can be grouped into three categories: domain experience, technical expertise, and visualization literacy.

\paragraph{Domain Experience}
Domain experience is most commonly reported in our corpus.
Forty-five studies included domain experience as an aspect of describing their participants.
We found that the researchers had different focus areas and notions when measuring and reporting the domain experience of their participants.
The knowledge about the domain can be assessed in certain domains with standardized questionnaires.
\citet{paper35} and \citet{paper10}, for example, both assess the musical sophistication of their participants measured by the Musical Sophistication Index~\cite{mullensiefen_musicality_2014}.
Others assess the domain knowledge of their participants by having them report their years of working in the field~\cite{paper34, paper85} or self-assess their experience on a scale~\cite{paper20, paper81}.
The level of the participants' experience is also measured by asking the participants how frequently they interact with the domain.
This is often the case in everyday-life recommendation scenarios such as movie or music streaming~\cite{paper120, paper93, paper89, paper39} or e-commerce~\cite{paper106, paper118}.
Overall, the papers reported that their participants are familiar with the evaluated domains and interact with them regularly.
This does not necessarily apply to the familiarity with a specific application. 
\citet{paper102} reported no participant had prior experience with the application.

\paragraph{Technical Expertise}
The participants' technical expertise was assessed in 13 studies.
The majority of the studies reported self-assessed technical expertise on varying scales.
We additionally found that different aspects of the technical expertise were evaluated.
\citet{paper24} and \citet{paper34} asked their participants about their experience with AI or Machine Learning and reported a general familiarity with AI.
\citet{paper39} and \citet{paper95} specifically asked about familiarity with recommender systems, while \citet{paper89} combined both aspects and recorded if their participants have taken courses in deep learning, information retrieval, or recommender systems.
These papers also reported a general familiarity with the aspects in question.
\citet{paper43, paper103} asked about the computer literacy of their participants. 
They reported that 90\% of their participants have a high to very high computer literacy.
Five studies did not provide information on the distribution of their participants regarding technical expertise.

\paragraph{Visualization Literacy}
The visualization literacy of the participants was recorded in eight studies.
However, only \citet{paper74} reported the distribution of their participants between low (21 participants) and high (51 participants) visualization literacy.

\subsection{The Impact of User Characteristics}
\label{sec:charactereval}

We analyzed the findings reported in our corpus to investigate which user characteristics introduced in Section~\ref{sec:participants} might impact the effect of explanations.
Overall, we extracted 165 findings from 31 papers in which the effect of explanations was evaluated on different user characteristics.
As seen in Table~\ref{tab:recorded-characteristics}, only a fraction of the papers that record a characteristic of their participants also report measurements on how the effect of explanations differs when disaggregating the participant data.

\begin{table}
\caption{Number of results and publications in which the effect of user characteristics on explanation effects was analyzed with the \textbf{explanation type} and \textbf{explanation generation method} as context variables. Rows with \checkmark indicate that an impact of the user characteristic was found; rows with $\times$ indicate that no impact was found. The last column sums the number of findings for each user characteristic. The last row sums up the findings where an effect was found or not found. 
The superscript letter indicates the explanation type: $t$: textual, $v$: visual, $h$: hybrid, $s$: schematic (i.e., a table), and $n$: numerical. Findings can contain multiple explanations of different types that were grouped together. 
Subscript letters show the explanation method: $p$: post-hoc, $i$: intrinsic. 
Articles where information on any of the categories was not provided or the recommendations were simulated are marked with $-$.}
\label{tab:charactereffects}
\Description{The first two columns list the user characteristics that were analyzed in the evaluations of our corpus sorted by their category (demographic, personality, experience).
Each characteristic consists of two rows, one in which the papers that found an effect of the user characteristic are referenced and one in which the papers that did not find an effect are referenced.
The superscript letter indicates the explanation type: t: textual, v: visual, h: hybrid, s: schematic (i.e., a table), and n: numerical. Note that one finding can be about multiple explanations of different types.}
\resizebox{\columnwidth}{!}{%
\begin{tabular}{@{}clllllllllr@{}}
\toprule
\multicolumn{1}{c}{} & \begin{tabular}[c]{@{}c@{}}User\\ Characteristic\end{tabular} & \begin{tabular}[c]{@{}c@{}}Effect \\ Found\end{tabular} & \multicolumn{1}{c}{Effectiveness} & \multicolumn{1}{c}{Efficiency} & \multicolumn{1}{c}{\begin{tabular}[c]{@{}c@{}}Perceived Explanation\\ Quality\end{tabular}} & \multicolumn{1}{c}{Persuasiveness} & \multicolumn{1}{c}{Transparency} & \multicolumn{1}{c}{Trust} & \multicolumn{1}{c}{Usability/UX} & \multicolumn{1}{c}{Total} \\ \midrule
\multirow{10}{*}{\rotatebox{90}{Demographic}} & \multirow{2}{*}{Age} & \checkmark &  &  & \cite{paper58}$^t_p$ & \cite{paper21}$^h_-$ &  &  &  & \multirow{2}{*}{2} \\ 
 &  & $\times$  &  &  &  &  &  &  & &  \\ \cmidrule{2-11}
 & \multirow{2}{*}{Location} & \checkmark &  &  &  &  & \cite{paper103}$^{ht}_p$ & \cite{paper103}$^{ht}_p$ &  & \multirow{2}{*}{6} \\ 
 &  & $\times$  &  &  &  &  &  &  &  &  \\ \cmidrule{2-11}
 & \multirow{2}{*}{Gender} & \checkmark &  &  & \cite{paper60}$^t_p$ &  & \cite{paper29}$^h_i$ & \cite{paper29}$^h_i$ & \cite{paper29}$^h_i$, \cite{paper99}$^t_-$ & \multirow{2}{*}{9} \\ \addlinespace
 &  & $\times$  &  &  & \cite{paper58}$^t_p$, \cite{paper97}$^h_-$ & \cite{paper21}$^h_-$ &  &  &  &  \\ \cmidrule{2-11}
 & \multirow{2}{*}{Education} & \checkmark &  &  &  &  & \cite{paper41}$^t_-$ &  & \cite{paper29}$^h_i$ & \multirow{2}{*}{6} \\ \addlinespace
 &  & $\times$  &  &  &  &  & \cite{paper29}$^h_i$, \cite{paper41}$^t_-$ & \cite{paper29}$^h_i$ & &  \\ \midrule 
\multirow{32}{*}{\rotatebox{90}{Personality}} & \multirow{2}{*}{Agreeableness} & \checkmark &  &  & \cite{paper27}$^t_-$ &  &  &  &  & \multirow{2}{*}{4} \\ \addlinespace
 &  & $\times$  &  &  & \cite{paper65}$^t_-$ & \cite{paper22}$^h_-$ &  &  &  &  \\ \cmidrule{2-11}
 & \multirow{2}{*}{\begin{tabular}[c]{@{}c@{}}Conscientious-\\ness\end{tabular}} & \checkmark &  &  & \cite{paper87}$^{tv}_i$ & \cite{paper21, paper22}$^h_-$ &  &  &  & \multirow{2}{*}{4} \\  \addlinespace
 &  & $\times$  &  &  &  & \cite{paper22}$^h_-$ &  &  &  &  \\ \cmidrule{2-11}
 & \multirow{2}{*}{\begin{tabular}[c]{@{}c@{}}Decision-Making\\ Type\end{tabular}} & \checkmark & \cite{paper69}$^{stv}_p$, \cite{paper75}$^{hv}_p$ & \cite{paper75}$^{hv}_p$ & \cite{paper69}$^{stv}_p$, \cite{paper75}$^{hv}_p$, \cite{paper97}$^h_-$ &  & \cite{paper69}$^{stv}_p$ & \cite{paper69}$^{stv}_p$, \cite{paper75}$^{hv}_p$ & \cite{paper69}$^v_p$ & \multirow{2}{*}{17} \\ \addlinespace
 &  & $\times$  &  &  & \cite{paper97}$^h_-$ &  & \cite{paper75}$^{hv}_p$ &  &  &  \\ \cmidrule{2-11}
 & \multirow{2}{*}{Extraversion} & \checkmark &  &  &  & \cite{paper21}$^h_-$ &  &  &  & \multirow{2}{*}{4} \\ \addlinespace
 &  & $\times$  &  &  & \cite{paper65}$^t_-$ & \cite{paper22}$^h_-$ &  &  &  &  \\ \cmidrule{2-11}
 & \multirow{2}{*}{\begin{tabular}[c]{@{}c@{}}Need for\\ Cognition\end{tabular}} & \checkmark &  &  & \cite{paper10}$^h_p$ &  & \cite{paper12}$^t_-$ &  &  & \multirow{2}{*}{17} \\ \addlinespace
 &  & $\times$  & \cite{paper18}$^{hv}_-$, \cite{paper23}$^{h}_-$ & \cite{paper18}$^{hv}_-$, \cite{paper23}$^{h}_-$ & \cite{paper57}$^t_-$, \cite{paper84}$^v_-$ & \cite{paper18}$^{hv}_-$, \cite{paper23}$^{h}_-$ & \cite{paper12}$^t_-$, \cite{paper18}$^{hv}_-$, \cite{paper23}$^{h}_-$ & \cite{paper18}$^{hv}_-$, \cite{paper23}$^{h}_-$ & \cite{paper18}$^{hv}_-$, \cite{paper23}$^{h}_-$ &  \\ \cmidrule{2-11}
 & \multirow{2}{*}{Neuroticism} & \checkmark &  &  &  & \cite{paper21}$^h_-$,\cite{paper87}$^t_i$ &  &  &  & \multirow{2}{*}{7} \\ \addlinespace
 &  & $\times$  &  &  & \cite{paper27, paper65}$^t_-$ & \cite{paper22}$^h_-$ &  &  &  &  \\ \cmidrule{2-11}
 & \multirow{2}{*}{Openness} & \checkmark &  &  & \cite{paper87}$^{tv}_i$ &  &  &  &  & \multirow{2}{*}{3} \\ 
 &  & $\times$  &  &  &  & \cite{paper22}$^h_-$ &  &  &  &  \\ \cmidrule{2-11}
 & \multirow{2}{*}{\begin{tabular}[c]{@{}c@{}}Personal\\ Innovativeness\end{tabular}} & \checkmark & \cite{paper18}$^{hv}_-$& \cite{paper18}$^{hv}_-$ &  &  & \cite{paper23}$^{h}_-$ &  & \cite{paper18}$^{hv}_-$ & \multirow{2}{*}{13} \\ \addlinespace
 &  & $\times$  & \cite{paper23}$^{h}_-$ & \cite{paper23}$^{h}_-$ &  & \cite{paper18}$^{hv}_-$, \cite{paper23}$^{h}_-$ & \cite{paper18}$^{hv}_-$ & \cite{paper18}$^{hv}_-$, \cite{paper23}$^{h}_-$ & \cite{paper18}$^{hv}_-$, \cite{paper23}$^{h}_-$ &  \\ \cmidrule{2-11}
 & \multirow{2}{*}{\begin{tabular}[c]{@{}c@{}}Trust\\ Propensity\end{tabular}} & \checkmark &  &  &  &  &  & \cite{paper21}$^h_-$ & \cite{paper23}$^{h}_-$ & \multirow{2}{*}{8} \\ \addlinespace
 &  & $\times$  & \cite{paper23}$^{h}_-$ & \cite{paper23}$^{h}_-$ &  & \cite{paper23}$^{h}_-$ & \cite{paper23}$^{h}_-$ & \cite{paper23}$^{h}_-$ & \cite{paper23}$^{h}_-$ &  \\ \cmidrule{2-11}
 & \multirow{2}{*}{Rationality} & \checkmark &  &  &  &  &  &  &  & \multirow{2}{*}{3} \\ 
 &  & $\times$  &  &  & \cite{paper68}$^t_-$ &  &  & \cite{paper107}$^t_-$ &  &  \\ \cmidrule{2-11}
 & \multirow{2}{*}{\begin{tabular}[c]{@{}c@{}}Social\\ Awareness\end{tabular}} & \checkmark & \cite{paper69}$^{stv}_p$, \cite{paper75}$^{hv}_p$ &  & \cite{paper69}$^{stv}_p$, \cite{paper75}$^{hv}_p$ & \cite{paper69}$^{stv}_p$ & \cite{paper36, paper69}$^{stv}_p$, \cite{paper75}$^{hn}_p$ & \cite{paper68}$^t_-$, \cite{paper69}$^{stv}_p$, \cite{paper75}$^{hv}_p$ &  & \multirow{2}{*}{15} \\ \addlinespace
 &  & $\times$  &  & \cite{paper75}$^{hv}_p$ & \cite{paper68}$^t_-$ &  &  &  &  &  \\ \cmidrule{2-11}
 & \multirow{2}{*}{\begin{tabular}[c]{@{}c@{}}Trust in\\ Technology\end{tabular}} & \checkmark &  &  &  &  & \cite{paper108}$^v_i$ &  &  & \multirow{2}{*}{1} \\ 
 &  & $\times$  &  &  &  &  &  &  &  &  \\ \cmidrule{2-11}
 & \multirow{2}{*}{Valence} & \checkmark &  &  & \cite{paper88}$^v_p$ &  &  &  &  & \multirow{2}{*}{1} \\ 
 &  & $\times$  &  &  &  &  &  &  &  &  \\ 
 \midrule
 \multirow{10}{*}{\rotatebox{90}{Experience}} & \multirow{2}{*}{\begin{tabular}[c]{@{}c@{}}Domain\\ Experience\end{tabular}} & \checkmark &  &  &  &  & \cite{paper108}$^v_i$ &  &  & \multirow{2}{*}{11} \\ \addlinespace
 &  & $\times$  & \cite{paper23}$^{h}_-$& \cite{paper23}$^{h}_-$ & \cite{paper87}$^{tv}_i$ & \cite{paper23}$^{h}_-$, \cite{paper87}$^{tv}_i$ & \cite{paper23}$^{h}_-$ & \cite{paper23}$^{h}_-$ & \cite{paper23}$^{h}_-$ &  \\
 \cmidrule{2-11}
 & \multirow{2}{*}{\begin{tabular}[c]{@{}c@{}}Technical\\ Expertise\end{tabular}} & \checkmark & \cite{paper34}$^{tv}_-$  & \cite{paper18}$^{hv}_-$&  & \cite{paper87}$^{t}_i$ &  &  &  & \multirow{2}{*}{16} \\ \addlinespace
 &  & $\times$  & \cite{paper18}$^{hv}_-$, \cite{paper23}$^{h}_-$ & \cite{paper23}$^{h}_-$ & \cite{paper87}$^{tv}_i$  & \cite{paper18}$^{hv}_-$, \cite{paper23}$^{h}_-$ & \cite{paper18}$^{hv}_-$, \cite{paper23}$^{h}_-$ & \cite{paper18}$^{hv}_-$, \cite{paper23}$^{h}_-$ & \cite{paper18}$^{hv}_-$, \cite{paper23}$^{h}_-$ &  \\ \cmidrule{2-11}
 & \multirow{2}{*}{\begin{tabular}[c]{@{}c@{}}Visualization\\ Literacy\end{tabular}} & \checkmark & \cite{paper23}$^{h}_-$ & \cite{paper23}$^{h}_-$ & \cite{paper75}$^{v}_p$ ,\cite{paper87}$^{v}_i$ &  & \cite{paper23}$^{ht}_-$ & \cite{paper23}$^{h}_-$ & \cite{paper23}$^{h}_-$, \cite{paper69}$^{stv}_p$ & \multirow{2}{*}{18} \\ \addlinespace
 &  & $\times$  & \cite{paper18}$^{hv}_-$ & \cite{paper18}$^{hv}_-$& \cite{paper87}$^{v}_i$ & \cite{paper18}$^{hv}_-$, \cite{paper23}$^{h}_-$ & \cite{paper18}$^{hv}_-$ & \cite{paper18}$^{hv}_-$ & \cite{paper18}$^{hv}_-$, \cite{paper23}$^{h}_-$ &  \\ 
 \midrule
\multicolumn{2}{c}{\multirow{2}{*}{\begin{tabular}[c]{@{}c@{}}Total\\ (number of results)\end{tabular}}} & \checkmark & \multicolumn{1}{c}{7} & \multicolumn{1}{c}{4} & \multicolumn{1}{c}{21} & \multicolumn{1}{c}{9} & \multicolumn{1}{c}{13} & \multicolumn{1}{c}{12} & \multicolumn{1}{c}{10} &  \\
\multicolumn{2}{c}{} & $\times$ & \multicolumn{1}{c}{8} & \multicolumn{1}{c}{8} & \multicolumn{1}{c}{18} & \multicolumn{1}{c}{18} & \multicolumn{1}{c}{13} & \multicolumn{1}{c}{12} & \multicolumn{1}{c}{14} &  \\ 
\bottomrule
\end{tabular}
}
\end{table}

We present these results in Tables \ref{tab:charactereffects}, \ref{tab:charactereffects_domain}, and \ref{tab:charactereffects_rs} with different context variables\footnote{An interactive version of these tables is available at \url{https://kathriwa.github.io/interactive-survey-visualization/\#/characteristics}}.
More precisely, we present the evaluated explanation effects on different user characteristics using as context variable(s) the evaluated explanation type and explanation method in Table \ref{tab:charactereffects}, the application domain in Table \ref{tab:charactereffects_domain}, and the type of recommender systems in Table \ref{tab:charactereffects_rs}.
At first glance, the data is sparse for all combinations of effects and characteristics, making it impossible to determine a statistically significant impact of user characteristics on explanation effects.
We can identify some indications that might lead to the generation of hypotheses that can be tested in further experiments.

\subsubsection{Demographics}
Notably, the demographic information, the category recorded most frequently in Section \ref{sec:result-dems}, is only evaluated by a minority of the papers.
The possible impacts of demographic characteristics on the effectiveness and efficiency of explanations were rarely evaluated.

We extracted nine findings from six papers that analyzed whether the participants' gender influenced the explanation effects.
Differences between genders were found in user experience, transparency, and trust.
No impact was found on persuasiveness, and perceived explanation quality has a slight majority of findings in which no effect was found. 
Due to the low number of studies, these can be merely seen as indications.

The number of findings and studies is even lower for the other user characteristics.
Location, education, and age were only evaluated by 1-2 studies.
\citet{paper103} assessed the impact of their participants' location on transparency and trust.
We extracted four findings that indicate that the location impacts trust for participants from France, Japan, and the USA.
At least for Japanese participants, this also seems to be the case for transparency.
Two papers looked at the impact of the education level on transparency, trust, and usability/UX perception.
\citet{paper29} compared graduate with undergraduate students and found higher levels of transparency, trust, and usability/UX perception in undergraduate students.
\citet{paper41} compared students with instructors and found mixed results on transparency.
Regarding age, \citet{paper21} found a correlation with persuasiveness, and \citet{paper58} found that younger participants perceive the explanation quality as higher than older participants.

\subsubsection{Personality}
The impact of personality traits on explanation effects was evaluated by 18 papers from which we could extract 107 findings.
In contrast to the demographics, the difference between the number of papers in which the characteristic was recorded and the number of papers in which it was evaluated is much smaller.

Most papers and results in the personality category were extracted for NFC.
Six papers reported 17 results in which the impact of NFC was evaluated on all seven effects stated in Table~\ref{tab:charactereffects}.
There seems to be a general tendency that no impact can be found, except for transparency and perceived explanation quality, where the results were mixed.

The Big Five traits were analyzed solely for a possible impact on perceived explanation quality and persuasiveness.
Neuroticism was evaluated most frequently.
Regarding persuasiveness, two papers found an impact of neuroticism; one did not.
Both papers that looked into the impact of neuroticism on the perceived explanation quality did not find statistically significant results.
The results for extraversion were similar; no impact was found on perceived explanation quality, but for persuasiveness, one paper found an impact while another did not.
\citet{paper87} indicated that conscientiousness and openness influence the perceived explanation quality.
However, it was the only paper in our corpus analyzing this combination of variables.
Regarding persuasiveness, no impact was found for openness, while the analyses of conscientiousness returned mixed results.
Only one of three studies found an impact of agreeableness on perceived explanation quality.
Only one study evaluated the impact of agreeableness on persuasiveness without finding evidence.

Five studies examined the effect of social awareness on transparency, trust, effectiveness, efficiency, persuasiveness, and perceived explanation quality.
All studies evaluating transparency, trust, effectiveness, and persuasiveness found that the results differ depending on the participants' social awareness scores.
The one study that evaluated a possible interaction of social awareness and efficiency did not find a significant effect~\cite{paper75}.
The impact of social awareness on perceived explanation quality was found in four studies~\cite{paper69, paper75} but not in one study~\cite{paper68}.
Personal innovativeness was analyzed by two papers~\cite{paper18, paper23}.
Both results contradicted for effectiveness, efficiency, transparency, and usability/UX.
Both papers did not find an impact on trust and persuasiveness. 
The two papers that evaluated the propensity to trust others~\cite{paper21, paper23} did not find an impact on effectiveness, efficiency, persuasiveness, and transparency.
The results were contradictory for trust and usability/UX.
No effect was found for rationality on perceived explanation quality and trust.
The only evaluation of trust in technology found that it can impact the transparency perceived by the participants~\cite{paper108}.
One study~\cite{paper88} found an impact of valence on perceived explanation quality.
The potential impact of the decision-making strategy was evaluated in three papers~\cite{paper69, paper75, paper97} on all effects but persuasiveness.
\citet{paper69, paper75} found impacts on usability/UX perception, effectiveness, efficiency, and trust. 
Perceived explanation quality and transparency yielded mixed results. 

\subsubsection{Experience}
The prior experience of the participants was recorded by 61 studies out of which 17 analyzed whether it impacts the explanation effect being evaluated.
The difference in the number of papers recording the characteristic and analyzing it is lower for the experience category than in the demographic but higher than in the personality category.

Seven studies evaluated if the domain experience of their participants had an impact on all of the explanation effects listed in Table~\ref{tab:recorded-characteristics}.
Contrasting results were found for transparency; evaluations for all other effects did not find significant results.
Four studies evaluated the impact of technical expertise on the explanation effect.
There was no impact on perceived explanation quality, transparency, trust, and usability/UX.
Some studies found an impact of technical expertise on effectiveness, efficiency, and persuasiveness; others did not.
Six studies looked into the impact of visualization literacy on the explanation effects.
No significant impact was found for persuasiveness; all other six effects had studies in which an impact was found and those in which it was not.

\subsubsection{Explanation Context}
We looked at four additional variables that provide context to the results: explanation type, explanation generation method, recommender type, and the application domain in which the explanation was evaluated.
In terms of explanation type, we found 68 results in which multiple explanations of different types were combined in the result. We cannot determine any patterns in instances where the result addresses only one explanation.
We often see that an impact of a user characteristic was found and not found for the same explanation type (see the trust or usability/UX effects for the trust propensity user characteristic in Table~\ref{tab:charactereffects} for an example), but no explanation type seems to be more likely to evoke an effect.
The explanation generation method is predominantly either unknown or the explanations were simulated for the user evaluation. 103 out of 165 findings are based on these explanations.
Intrinsically explainable recommender systems are rare in this evaluation. Only 21 of the findings stem from a model intrinsic explanation such that a comparison between intrinsic and post-hoc explanations is not possible.
%
\begin{table}
\caption{Number of results and publications in which the effect of user characteristics on explanation effects was analyzed with the \textbf{application domain} as a context variable. Rows with \checkmark indicate that an impact of the user characteristic was found; rows with $\times$ indicate that no impact was found. The last column sums the number of findings for each user characteristic. The last row sums up the findings where an effect was found or not found. 
Application domains are depicted as follows: 
\doc: document, \ecom: e-commerce, \education: education, \energy: energy saving, \health: health, \movie: movie, \music: music, \dating: online dating, \poi: point-of-interest, \social: social.
}
\label{tab:charactereffects_domain}
\Description{The first two columns list the user characteristics that were analyzed in the evaluations of our corpus sorted by their category (demographic, personality, experience).
Each characteristic consists of two rows, one in which the papers that found an effect of the user characteristic are referenced and one in which the papers that did not find an effect are referenced.}
\resizebox{\columnwidth}{!}{%
\begin{tabular}{@{}clllllllllr@{}}
\toprule
\multicolumn{1}{c}{} & \begin{tabular}[c]{@{}c@{}}User\\ Characteristic\end{tabular} & \begin{tabular}[c]{@{}c@{}}Effect \\ Found\end{tabular} & \multicolumn{1}{c}{Effectiveness} & \multicolumn{1}{c}{Efficiency} & \multicolumn{1}{c}{\begin{tabular}[c]{@{}c@{}}Perceived Explanation\\ Quality\end{tabular}} & \multicolumn{1}{c}{Persuasiveness} & \multicolumn{1}{c}{Transparency} & \multicolumn{1}{c}{Trust} & \multicolumn{1}{c}{Usability/UX} & \multicolumn{1}{c}{Total} \\ \midrule
\multirow{10}{*}{\rotatebox{90}{Demographic}} & \multirow{2}{*}{Age} & \checkmark &  &  & \cite{paper58} \movie & \cite{paper21} \health &  &  &  & \multirow{2}{*}{2} \\ 
 &  & $\times$  &  &  &  &  &  &  & &  \\ \cmidrule{2-11}
 & \multirow{2}{*}{Location} & \checkmark &  &  &  &  & \cite{paper103} \movie & \cite{paper103} \movie &  & \multirow{2}{*}{6} \\ 
 &  & $\times$  &  &  &  &  &  &  &  &  \\ \cmidrule{2-11}
 & \multirow{2}{*}{Gender} & \checkmark &  &  & \cite{paper60} \dating &  & \cite{paper29} \education & \cite{paper29} \education & \cite{paper29} \education, \cite{paper99} \poi & \multirow{2}{*}{9} \\ \addlinespace
 &  & $\times$  &  &  & \cite{paper58} \movie, \cite{paper97} \poi & \cite{paper21} \health &  &  &  &  \\ \cmidrule{2-11}
 & \multirow{2}{*}{Education} & \checkmark &  &  &  &  & \cite{paper41} \education &  & \cite{paper29} \education & \multirow{2}{*}{6} \\ \addlinespace
 &  & $\times$  &  &  &  &  & \cite{paper29, paper41} \education & \cite{paper29} \education & &  \\ \midrule 
\multirow{32}{*}{\rotatebox{90}{Personality}} & \multirow{2}{*}{Agreeableness} & \checkmark &  &  & \cite{paper27} \poi &  &  &  &  & \multirow{2}{*}{4} \\ \addlinespace
 &  & $\times$  &  &  & \cite{paper65} \poi & \cite{paper22} \health &  &  &  &  \\ \cmidrule{2-11}
 & \multirow{2}{*}{\begin{tabular}[c]{@{}c@{}}Conscientious-\\ness\end{tabular}} & \checkmark &  &  & \cite{paper87} \music & \cite{paper21, paper22} \health &  &  &  & \multirow{2}{*}{4} \\  \addlinespace
 &  & $\times$  &  &  &  & \cite{paper22} \health &  &  &  &  \\ \cmidrule{2-11}
 & \multirow{2}{*}{\begin{tabular}[c]{@{}c@{}}Decision-Making\\ Type\end{tabular}} & \checkmark & \cite{paper69, paper75} \poi & \cite{paper75} \poi & \cite{paper69, paper75, paper97} \poi &  & \cite{paper69} \poi & \cite{paper69, paper75} \poi & \cite{paper69} \poi & \multirow{2}{*}{17} \\ \addlinespace
 &  & $\times$  &  &  & \cite{paper97} \poi &  & \cite{paper75} \poi &  &  &  \\ \cmidrule{2-11}
 & \multirow{2}{*}{Extraversion} & \checkmark &  &  &  & \cite{paper21} \health &  &  &  & \multirow{2}{*}{4} \\ \addlinespace
 &  & $\times$  &  &  & \cite{paper65} \poi & \cite{paper22} \health &  &  &  &  \\ \cmidrule{2-11}
 & \multirow{2}{*}{\begin{tabular}[c]{@{}c@{}}Need for\\ Cognition\end{tabular}} & \checkmark &  &  & \cite{paper10} \music &  & \cite{paper12} \health &  &  & \multirow{2}{*}{17} \\ \addlinespace
 &  & $\times$  & \cite{paper18} \social, \cite{paper23} \doc & \cite{paper18} \social, \cite{paper23} \doc & \cite{paper57} \ecom, \cite{paper84} \music & \cite{paper18} \social, \cite{paper23} \doc & \cite{paper12} \health, \cite{paper18} \social, \cite{paper23} \doc & \cite{paper18} \social, \cite{paper23} \doc & \cite{paper18} \social, \cite{paper23} \doc &  \\ \cmidrule{2-11}
 & \multirow{2}{*}{Neuroticism} & \checkmark &  &  &  & \cite{paper21} \health,\cite{paper87} \music &  &  &  & \multirow{2}{*}{7} \\ \addlinespace
 &  & $\times$  &  &  & \cite{paper27, paper65} \poi & \cite{paper22} \health &  &  &  &  \\ \cmidrule{2-11}
 & \multirow{2}{*}{Openness} & \checkmark &  &  & \cite{paper87} \music &  &  &  &  & \multirow{2}{*}{3} \\ 
 &  & $\times$  &  &  &  & \cite{paper22} \health &  &  &  &  \\ \cmidrule{2-11}
 & \multirow{2}{*}{\begin{tabular}[c]{@{}c@{}}Personal\\ Innovativeness\end{tabular}} & \checkmark & \cite{paper18} \social& \cite{paper18} \social &  &  & \cite{paper23} \doc &  & \cite{paper18} \social & \multirow{2}{*}{13} \\ \addlinespace
 &  & $\times$  & \cite{paper23} \doc & \cite{paper23} \doc &  & \cite{paper18} \social, \cite{paper23} \doc & \cite{paper18} \social & \cite{paper18} \social, \cite{paper23} \doc & \cite{paper18} \social, \cite{paper23} \doc &  \\ \cmidrule{2-11}
 & \multirow{2}{*}{\begin{tabular}[c]{@{}c@{}}Trust\\ Propensity\end{tabular}} & \checkmark &  &  &  &  &  & \cite{paper21} \health & \cite{paper23} \doc & \multirow{2}{*}{8} \\ \addlinespace
 &  & $\times$  & \cite{paper23} \doc & \cite{paper23} \doc &  & \cite{paper23} \doc & \cite{paper23} \doc & \cite{paper23} \doc & \cite{paper23} \doc &  \\ \cmidrule{2-11}
 & \multirow{2}{*}{Rationality} & \checkmark &  &  &  &  &  &  &  & \multirow{2}{*}{3} \\ 
 &  & $\times$  &  &  & \cite{paper68} \poi &  &  & \cite{paper107} \ecom &  &  \\ \cmidrule{2-11}
 & \multirow{2}{*}{\begin{tabular}[c]{@{}c@{}}Social\\ Awareness\end{tabular}} & \checkmark & \cite{paper69, paper75} \poi &  & \cite{paper69, paper75} \poi & \cite{paper69} \poi & \cite{paper36} \energy, \cite{paper69, paper75} \poi & \cite{paper68, paper69, paper75} \poi &  & \multirow{2}{*}{15} \\ \addlinespace
 &  & $\times$  &  & \cite{paper75} \poi & \cite{paper68} \poi &  &  &  &  &  \\ \cmidrule{2-11}
 & \multirow{2}{*}{\begin{tabular}[c]{@{}c@{}}Trust in\\ Technology\end{tabular}} & \checkmark &  &  &  &  & \cite{paper108} \movie &  &  & \multirow{2}{*}{1} \\ 
 &  & $\times$  &  &  &  &  &  &  &  &  \\ \cmidrule{2-11}
 & \multirow{2}{*}{Valence} & \checkmark &  &  & \cite{paper88} \movie &  &  &  &  & \multirow{2}{*}{1} \\ 
 &  & $\times$  &  &  &  &  &  &  &  &  \\ 
 \midrule
 \multirow{10}{*}{\rotatebox{90}{Experience}} & \multirow{2}{*}{\begin{tabular}[c]{@{}c@{}}Domain\\ Experience\end{tabular}} & \checkmark &  &  &  &  & \cite{paper108} \movie &  &  & \multirow{2}{*}{11} \\ \addlinespace
 &  & $\times$  & \cite{paper23} \doc & \cite{paper23} \doc & \cite{paper87} \music & \cite{paper23} \doc, \cite{paper87} \music & \cite{paper23} \doc & \cite{paper23} \doc & \cite{paper23} \doc &  \\
 \cmidrule{2-11}
 & \multirow{2}{*}{\begin{tabular}[c]{@{}c@{}}Technical\\ Expertise\end{tabular}} & \checkmark & \cite{paper34} \health  & \cite{paper18} \social &  & \cite{paper87} \music &  &  &  & \multirow{2}{*}{16} \\ \addlinespace
 &  & $\times$  & \cite{paper18} \social, \cite{paper23} \doc & \cite{paper23} \doc & \cite{paper87} \music & \cite{paper18} \social, \cite{paper23} \doc & \cite{paper18} \social, \cite{paper23} \doc & \cite{paper18} \social, \cite{paper23} \doc & \cite{paper18} \social, \cite{paper23} \doc &  \\ \cmidrule{2-11}
 & \multirow{2}{*}{\begin{tabular}[c]{@{}c@{}}Visualization\\ Literacy\end{tabular}} & \checkmark & \cite{paper23} \doc & \cite{paper23} \doc & \cite{paper75} \poi ,\cite{paper87} \music &  & \cite{paper23} \doc & \cite{paper23} \doc & \cite{paper23} \doc, \cite{paper69} \poi & \multirow{2}{*}{18} \\ \addlinespace
 &  & $\times$  & \cite{paper18} \social & \cite{paper18} \social & \cite{paper87} \music & \cite{paper18} \social, \cite{paper23} \doc & \cite{paper18} \social & \cite{paper18} \social & \cite{paper18} \social, \cite{paper23} \doc &  \\ 
 \midrule
\multicolumn{2}{c}{\multirow{2}{*}{\begin{tabular}[c]{@{}c@{}}Total\\ (number of results)\end{tabular}}} & \checkmark & \multicolumn{1}{c}{7} & \multicolumn{1}{c}{4} & \multicolumn{1}{c}{21} & \multicolumn{1}{c}{9} & \multicolumn{1}{c}{13} & \multicolumn{1}{c}{12} & \multicolumn{1}{c}{10} &  \\
\multicolumn{2}{c}{} & $\times$ & \multicolumn{1}{c}{8} & \multicolumn{1}{c}{8} & \multicolumn{1}{c}{18} & \multicolumn{1}{c}{18} & \multicolumn{1}{c}{13} & \multicolumn{1}{c}{12} & \multicolumn{1}{c}{14} &  \\ 
\bottomrule
\end{tabular}
}
\end{table}
Along with the explanation type and generation method, we investigated the results presented in Table~\ref{tab:charactereffects} in the context of the application domain in which the explanations were evaluated (Table~\ref{tab:charactereffects_domain}) and the type of recommender system that was explained (Table~\ref{tab:charactereffects_rs}).
We identified a total of nine different application domains that were evaluated, with point-of-interest recommendations being the most common (44 findings), followed by document (39 findings), and social recommendations (27 findings). We notice the dominance of some domains in certain user characteristics. The decision-making type and social awareness, for example, were almost exclusively evaluated in the point-of-interest domain and trust propensity and the education level of participants in the document and education domain, respectively.
%
\begin{table}
\caption{Number of results and publications in which the effect of user characteristics on explanation effects was analyzed and the \textbf{type of recommender system} as context variable. Rows with \checkmark indicate that an impact of the user characteristic was found; rows with $\times$ indicate that no impact was found. The last column sums the number of findings for each user characteristic. The last row sums up the findings where an effect was found or not found. 
The recommender type is shown after the domain and indicated by $cb$ for content-based, $cf$ for collaborative filtering, and $hy$ for hybrid systems. 
Articles where information on any of the categories was not provided or the recommendations were simulated are marked with $-$.}
\label{tab:charactereffects_rs}
\Description{The first two columns list the user characteristics that were analyzed in the evaluations of our corpus sorted by their category (demographic, personality, experience).
Each characteristic consists of two rows, one in which the papers that found an effect of the user characteristic are referenced and one in which the papers that did not find an effect are referenced.
The superscript letter indicates the explanation type: t: textual, v: visual, h: hybrid, s: schematic (i.e., a table), and n: numerical. Note that one finding can be about multiple explanations of different types.}
\resizebox{\columnwidth}{!}{%
\begin{tabular}{@{}clllllllllr@{}}
\toprule
\multicolumn{1}{c}{} & \begin{tabular}[c]{@{}c@{}}User\\ Characteristic\end{tabular} & \begin{tabular}[c]{@{}c@{}}Effect \\ Found\end{tabular} & \multicolumn{1}{c}{Effectiveness} & \multicolumn{1}{c}{Efficiency} & \multicolumn{1}{c}{\begin{tabular}[c]{@{}c@{}}Perceived Explanation\\ Quality\end{tabular}} & \multicolumn{1}{c}{Persuasiveness} & \multicolumn{1}{c}{Transparency} & \multicolumn{1}{c}{Trust} & \multicolumn{1}{c}{Usability/UX} & \multicolumn{1}{c}{Total} \\ \midrule
\multirow{10}{*}{\rotatebox{90}{Demographic}} & \multirow{2}{*}{Age} & \checkmark &  &  & \cite{paper58}$cb$ & \cite{paper21}$-$ &  &  &  & \multirow{2}{*}{2} \\ 
 &  & $\times$  &  &  &  &  &  &  & &  \\ \cmidrule{2-11}
 & \multirow{2}{*}{Location} & \checkmark &  &  &  &  & \cite{paper103}$-$ & \cite{paper103}$-$ &  & \multirow{2}{*}{6} \\ 
 &  & $\times$  &  &  &  &  &  &  &  &  \\ \cmidrule{2-11}
 & \multirow{2}{*}{Gender} & \checkmark &  &  & \cite{paper60}$cb$ &  & \cite{paper29}$cf$ & \cite{paper29}$cf$ & \cite{paper29}$cf$, \cite{paper99}$-$ & \multirow{2}{*}{9} \\ \addlinespace
 &  & $\times$  &  &  & \cite{paper58}$cb$, \cite{paper97}$-$ & \cite{paper21}$-$ &  &  &  &  \\ \cmidrule{2-11}
 & \multirow{2}{*}{Education} & \checkmark &  &  &  &  & \cite{paper41}$hy$ &  & \cite{paper29}$cf$ & \multirow{2}{*}{6} \\ \addlinespace
 &  & $\times$  &  &  &  &  & \cite{paper29}$cf$, \cite{paper41}$hy$ & \cite{paper29}$cf$ & &  \\ \midrule 
\multirow{32}{*}{\rotatebox{90}{Personality}} & \multirow{2}{*}{Agreeableness} & \checkmark &  &  & \cite{paper27}$-$ &  &  &  &  & \multirow{2}{*}{4} \\ \addlinespace
 &  & $\times$  &  &  & \cite{paper65}$-$ & \cite{paper22}$cb$ &  &  &  &  \\ \cmidrule{2-11}
 & \multirow{2}{*}{\begin{tabular}[c]{@{}c@{}}Conscientious-\\ness\end{tabular}} & \checkmark &  &  & \cite{paper87}$hy$ & \cite{paper21}$-$, \cite{paper22}$cb$ &  &  &  & \multirow{2}{*}{4} \\  \addlinespace
 &  & $\times$  &  &  &  & \cite{paper22}$cb$ &  &  &  &  \\ \cmidrule{2-11}
 & \multirow{2}{*}{\begin{tabular}[c]{@{}c@{}}Decision-Making\\ Type\end{tabular}} & \checkmark & \cite{paper69, paper75}$cb$ & \cite{paper75}$cb$ & \cite{paper69, paper75}$cb$, \cite{paper97}$-$ &  & \cite{paper69}$cb$ & \cite{paper69, paper75}$cb$ & \cite{paper69}$cb$ & \multirow{2}{*}{17} \\ \addlinespace
 &  & $\times$  &  &  & \cite{paper97}$-$ &  & \cite{paper75}$cb$ &  &  &  \\ \cmidrule{2-11}
 & \multirow{2}{*}{Extraversion} & \checkmark &  &  &  & \cite{paper21}$-$ &  &  &  & \multirow{2}{*}{4} \\ \addlinespace
 &  & $\times$  &  &  & \cite{paper65}$-$ & \cite{paper22}$cb$ &  &  &  &  \\ \cmidrule{2-11}
 & \multirow{2}{*}{\begin{tabular}[c]{@{}c@{}}Need for\\ Cognition\end{tabular}} & \checkmark &  &  & \cite{paper10}$hy$ &  & \cite{paper12}$-$ &  &  & \multirow{2}{*}{17} \\ \addlinespace
 &  & $\times$  & \cite{paper18, paper23}$cb$ & \cite{paper18, paper23}$cb$ & \cite{paper57}$cb cf$, \cite{paper84}$-$ & \cite{paper18, paper23}$cb$ & \cite{paper12}$-$, \cite{paper18, paper23}$cb$ & \cite{paper18, paper23}$cb$ & \cite{paper18, paper23}$cb$ &  \\ \cmidrule{2-11}
 & \multirow{2}{*}{Neuroticism} & \checkmark &  &  &  & \cite{paper21}$-$,\cite{paper87}$hy$ &  &  &  & \multirow{2}{*}{7} \\ \addlinespace
 &  & $\times$  &  &  & \cite{paper27, paper65}$-$ & \cite{paper22}$cb$ &  &  &  &  \\ \cmidrule{2-11}
 & \multirow{2}{*}{Openness} & \checkmark &  &  & \cite{paper87}$hy$ &  &  &  &  & \multirow{2}{*}{3} \\ 
 &  & $\times$  &  &  &  & \cite{paper22}$cb$ &  &  &  &  \\ \cmidrule{2-11}
 & \multirow{2}{*}{\begin{tabular}[c]{@{}c@{}}Personal\\ Innovativeness\end{tabular}} & \checkmark & \cite{paper18}$cb$ & \cite{paper18}$cb$ &  &  & \cite{paper23}$cb$ &  & \cite{paper18}$cb$ & \multirow{2}{*}{13} \\ \addlinespace
 &  & $\times$  & \cite{paper23}$cb$ & \cite{paper23}$cb$ &  & \cite{paper18, paper23}$cb$ & \cite{paper18}$cb$ & \cite{paper18, paper23}$cb$ & \cite{paper18, paper23}$cb$ &  \\ \cmidrule{2-11}
 & \multirow{2}{*}{\begin{tabular}[c]{@{}c@{}}Trust\\ Propensity\end{tabular}} & \checkmark &  &  &  &  &  & \cite{paper21}$-$ & \cite{paper23}$cb$ & \multirow{2}{*}{8} \\ \addlinespace
 &  & $\times$  & \cite{paper23}$cb$ & \cite{paper23}$cb$ &  & \cite{paper23}$cb$ & \cite{paper23}$cb$ & \cite{paper23}$cb$ & \cite{paper23}$cb$ &  \\ \cmidrule{2-11}
 & \multirow{2}{*}{Rationality} & \checkmark &  &  &  &  &  &  &  & \multirow{2}{*}{3} \\ 
 &  & $\times$  &  &  & \cite{paper68}$-$ &  &  & \cite{paper107}$-$ &  &  \\ \cmidrule{2-11}
 & \multirow{2}{*}{\begin{tabular}[c]{@{}c@{}}Social\\ Awareness\end{tabular}} & \checkmark & \cite{paper69, paper75}$cb$ &  & \cite{paper69, paper75}$cb$ & \cite{paper69}$cb$ & \cite{paper36}$-$, \cite{paper69, paper75}$cb$ & \cite{paper68}$-$, \cite{paper69, paper75}$cb$ &  & \multirow{2}{*}{15} \\ \addlinespace
 &  & $\times$  &  & \cite{paper75}$cb$ & \cite{paper68}$-$ &  &  &  &  &  \\ \cmidrule{2-11}
 & \multirow{2}{*}{\begin{tabular}[c]{@{}c@{}}Trust in\\ Technology\end{tabular}} & \checkmark &  &  &  &  & \cite{paper108}$cf$ &  &  & \multirow{2}{*}{1} \\ 
 &  & $\times$  &  &  &  &  &  &  &  &  \\ \cmidrule{2-11}
 & \multirow{2}{*}{Valence} & \checkmark &  &  & \cite{paper88}$cb$ &  &  &  &  & \multirow{2}{*}{1} \\ 
 &  & $\times$  &  &  &  &  &  &  &  &  \\ 
 \midrule
 \multirow{10}{*}{\rotatebox{90}{Experience}} & \multirow{2}{*}{\begin{tabular}[c]{@{}c@{}}Domain\\ Experience\end{tabular}} & \checkmark &  &  &  &  & \cite{paper108}$cf$ &  &  & \multirow{2}{*}{11} \\ \addlinespace
 &  & $\times$  & \cite{paper23}$cb$& \cite{paper23}$cb$ & \cite{paper87}$hy$ & \cite{paper23}$cb$, \cite{paper87}$hy$ & \cite{paper23}$cb$ & \cite{paper23}$cb$ & \cite{paper23}$cb$ &  \\
 \cmidrule{2-11}
 & \multirow{2}{*}{\begin{tabular}[c]{@{}c@{}}Technical\\ Expertise\end{tabular}} & \checkmark & \cite{paper34}$-$  & \cite{paper18}$cb$ &  & \cite{paper87}$hy$ &  &  &  & \multirow{2}{*}{16} \\ \addlinespace
 &  & $\times$  & \cite{paper18, paper23}$cb$ & \cite{paper23}$cb$ & \cite{paper87}$hy$  & \cite{paper18, paper23}$cb$ & \cite{paper18, paper23}$cb$ & \cite{paper18, paper23}$cb$ & \cite{paper18, paper23}$cb$ &  \\ \cmidrule{2-11}
 & \multirow{2}{*}{\begin{tabular}[c]{@{}c@{}}Visualization\\ Literacy\end{tabular}} & \checkmark & \cite{paper23}$cb$ & \cite{paper23}$cb$ & \cite{paper75}$cb$ ,\cite{paper87}$hy$ &  & \cite{paper23}$cb$ & \cite{paper23}$cb$ & \cite{paper23, paper69}$cb$ & \multirow{2}{*}{18} \\ \addlinespace
 &  & $\times$  & \cite{paper18}$cb$ & \cite{paper18}$cb$ & \cite{paper87}$hy$ & \cite{paper18, paper23}$cb$ & \cite{paper18}$cb$ & \cite{paper18}$cb$ & \cite{paper18, paper23}$cb$ &  \\ 
 \midrule
\multicolumn{2}{c}{\multirow{2}{*}{\begin{tabular}[c]{@{}c@{}}Total\\ (number of results)\end{tabular}}} & \checkmark & \multicolumn{1}{c}{7} & \multicolumn{1}{c}{4} & \multicolumn{1}{c}{21} & \multicolumn{1}{c}{9} & \multicolumn{1}{c}{13} & \multicolumn{1}{c}{12} & \multicolumn{1}{c}{10} &  \\
\multicolumn{2}{c}{} & $\times$ & \multicolumn{1}{c}{8} & \multicolumn{1}{c}{8} & \multicolumn{1}{c}{18} & \multicolumn{1}{c}{18} & \multicolumn{1}{c}{13} & \multicolumn{1}{c}{12} & \multicolumn{1}{c}{14} &  \\ 
\bottomrule
\end{tabular}
}
\end{table}
The dominant type of recommender system is content-based, with 107 out of 165 findings. Pure collaborative filtering systems are only used to generate the recommendations for 10 findings.

\section{Discussion}
\label{sec:discuss}

Overall, we observe that the data about the participants varies between the publications in terms of completeness.
Table \ref{tab:recorded-characteristics} illustrates that none of our analyzed characteristics were recorded by all papers in the corpus.
Reporting participants' age, gender, location, education, and prior domain experience seems standard practice. However, disaggregating the data by these features to see if the evaluated effect differs between sub-groups is rare.

\subsection{Who was recruited to evaluate recommender systems explanations?}
\label{sec:discuss-participant}

In this subsection, we discuss our results to answer \textbf{RQ 1}.
Our analyses demonstrate that it is challenging to aggregate the data in our corpus due to non-uniform ways of measuring and reporting it.
Therefore, we cannot conclusively determine an average participant for each application domain. To the best of our knowledge, there is no related work that investigated how representative study participants are in recommender systems evaluation, which leaves us without a reference point to determine whether the reported participant data in our corpus is sampled accordingly. Instead, we focus our discussion on the characteristics reported by at least one-quarter of the papers in our corpus.
We specifically discuss the participants' age, gender, location, education, and prior domain experience.

\paragraph{Age}
Regarding the participants' age, we made two main observations.
First, almost all participants are above the age of 18. 
One possible explanation for this might be age restrictions for crowd workers.
Our second observation is that the average age of the participants does not seem to reflect the average age of the overall population.
In our corpus, the age averages are predominantly between the ages of 20 and 40, while the average age of the overall population, especially in the countries where the participants were recruited, is shifting towards an older average.
A possible reason for this could be the recruiting method.
A large proportion of the papers in our corpus recruited their participants on crowdsourcing platforms. The findings of \citet{difallah2018demographics} support our observation, as they found that MTurk workers tend to be younger than the overall population.
In their study, conducted in 2018, 60\% of the workers were born after 1980, which also aligns with our findings.
Studies that did not rely on crowdsourcing platforms frequently recruited participants at universities. In such cases, students often comprised most of the sample.
This can be another explanation for the comparatively low age average.

\paragraph{Gender}
The difference between male and female participants was balanced in about half of the papers that reported the participant distribution for gender.
For the other half of the papers, a male majority was more common than a female majority.
We found very few papers report non-binary options in their gender distribution.
This does not necessarily imply that these options were not available in the study, but it could also mean that all participants identified as male or female, and the possibilities with no answer were omitted from the paper.
Especially with the demographic data, we rarely had access to the questions and response options given in the study.
As the exact questions are not provided, no conclusion can be drawn about the options provided.

\paragraph{Location}
The results for the participants' location show a clear dominance of participants living in the USA.
Again, the recruiting mechanism might impact the location as Amazon Mechanical Turk is a frequently used crowdsourcing platform in our corpus with a predominantly US population~\cite{difallah2018demographics}.

\paragraph{Education}
The results regarding the education level of the participants show that the majority of the participants evaluating recommendation explanations are highly educated.
Participants with at least a bachelor's degree were the dominant group in our sample.
The average participant has a higher education level than the average population. 

\paragraph{Domain Experience}
Regarding prior experience with the domain, the results show that most participants frequently interact with the application domain.
One reason might be that the evaluated domains are typically accessible to various people and predominant in the users' leisure time.
The most common are point-of-interest (e.g., restaurants or tourist attractions), e-commerce, movie, or music recommendations.
Another reason for the dominance of these domains might be their popularity in the recommender systems community.
\citet{chin_datasets_2022} analyzed popular datasets for recommender systems evaluation, which almost exclusively came from these domains.

The domain experience might also be related to the participant's age group, assuming that younger people who grew up with online shopping and media streaming services interact with these applications more frequently than the average above 60-year-old person.

\paragraph{Summary} In conclusion, the results of the frequently recorded characteristics show that the most common participant is an educated male or female in their twenties or thirties, living in the USA, and regularly interacting with everyday recommender systems.
Due to the low number of papers that reported the participant distribution for the remaining characteristics we analyzed (10\%  or less of the total corpus size), we cannot draw other observations about other demographic characteristics or their personality traits.
However, the characteristics we have provided enough information to match the WEIRD societies outlined in \citet{henrich_heine_norenzayan_2010}.

\subsection{What user groups were not represented?}
\label{sec:discuss-blind-spots}

After discussing what we know about the participants on which explanations are evaluated, we now focus on possible blind spots where user characteristics are underrepresented or missing.

Overall, for each of our analyzed characteristics, we found many papers that did not report recording it.
Therefore, we only refer to the papers in which the characteristic was reported when pointing out a gap.
We cannot completely rule out that these user groups were part of the participant sample of the papers that did not disclose the information.

Regarding participant diversity, the papers in our corpus did not include people from the Global South and Eastern Europe, a significant share of the population.
This issue has been pointed out in multiple publications such as \citet{Sturm_weirdHCI, Linxen_2021_weirdCHI} and is generally known in the HCI community. Given the culturally-dependent differences in user interaction \cite{Reinecke_culturally_adapted}, this may have a profound implication on the generalizability of the results to global settings.

Children, teenagers, and older people were rarely part of the participant sample.
People in these age groups might have different needs than younger adults regarding explanation content and display, so it would be interesting to see an evaluation of the same explanation that includes these user groups.
\citet{pmlr-v81-ekstrand18b}, for example, looked at the recommendation effectiveness and found that the recommendation accuracy differs when disaggregating the data by age and gender.
While one might argue that children only start independently interacting with most recommender systems applications from a certain age, this might change.
Furthermore, investigating explainable recommender systems in use cases where parents use them together with their children opens up a new research direction.

The comparisons regarding the participants' education level mainly investigate differences between participants with an above-average education level and participants with an even higher education level (e.g., undergraduate compared to graduate students as seen in \citet{paper29}).
Participants without an academic degree are in the minority in most studies.
Instead of comparing participant groups where the overall education level is high already, an evaluation of the effect of explanations on participants with an average or below-average education level would be interesting to see.

Regarding gender, we found only a few studies reporting non-binary participants.
As mentioned in Section \ref{sec:discuss-participant}, most papers did not report any gender other than male/female.

Only a small fraction of our corpus reported the personality traits of their participants.
However, we cannot identify user groups that might have been left out due to how these traits are reported.
In most cases, we do not know anything about the participant distribution.
We noticed, however, that the participants are frequently split into high and low-scoring, often based on a median split.

In terms of prior experience, we mainly found tech-savvy participants with domain experience in the studies of our corpus.
The technical expertise might be related to recruiting participants online, who are often experienced crowd workers, which requires a certain level of expertise.
However, in real life, many recommender systems applications also target less experienced users or might not know that an algorithm selects the items they see.
Although recruiting these participants might be more challenging, depending on the application domain, it could be relevant to find out how findings generalize to the general populations with all kinds of domain experience.

We do not have enough information about the participant distributions of the remaining characteristics to infer user groups that were not represented.
As of today, we do not know if these characteristics might impact the explanation effects, so we encourage reporting participant distributions and exploring possible impacts on explanation effects in future work.

\subsection{Which user characteristics impact the effect of explanations in recommender systems?}
\label{sec:discuss-charcteristic-impact}

In this subsection, we discuss the results of Section \ref{sec:charactereval} and answer \textbf{RQ 2}.
Generally, the number of findings we could extract in which the impact of user characteristics on the explanation effect was evaluated is very low.
In our corpus, we have at most five studies that evaluated the same characteristic and the same effect (impact of decision-making strategy on perceived explanation quality).
It is far more frequent that only one or two studies examined the same characteristic-effect combination. 
Therefore, it is not possible to derive significant results from our data.
Results, where multiple articles either all found or did not find an impact of a user characteristic on an effect, might indicate a tendency that can be formulated into hypotheses for further experimentation.
Table \ref{tab:charactereffects} shows the gaps the community should close and allows us to spot these indications easily.
Looking at \emph{social awareness}, for example, all three articles that investigated the impact on transparency and trust and both articles that looked into effectiveness found an effect.
This can be taken as motivation for further research that systematically evaluates the impact of social awareness on these explanation effects.

In the instances where an impact was observed in one study and not found in another, other factors than the user characteristics might have impacted the results.
If we look, for example, at the impact of extraversion on the persuasiveness of the explanations, \citet{paper21} found an effect while \citet{paper22} did not.
The work of \citet{paper21} replicates the study performed in \citet{paper22}, focusing on participants from India.
While their participants had a similar average age and education level, they were from different locations.
Additionally, the participant sample in \citet{paper21} had a male majority (39 male majority, 21 female, one other) while the participants in \citet{paper22} were predominantly female (21 male, 48 female).
Both of these aspects could explain the different results.

Most cases in which the explanation effects findings were inconclusive were independent studies evaluating different aspects.
Looking at the impact of domain knowledge on transparency, \citet{paper108} found an effect, and \citet{paper23} did not.
We do not know a lot about the participants in \citet{paper108} other than that they are mainly female students with an age average about ten years lower than the participants in the study by \citet{paper23}, who recruited researchers with at least one scientific publication.
Furthermore, both papers differ in the application domain, evaluated explanations, and research goals.
\citet{paper108} proposes a new explainable recommendation method, which they evaluate in the movie domain, while \citet{paper23} investigates different detail levels of explanations in the document recommendation domain.
Both studies evaluated different explanations with different modalities, complexities, and interaction features.
All these aspects could have led to the differences in results, but the differences could also stem from factors that have not been considered yet.

We also found, for example, ambiguous results within the same study.
This can occur when there are two different studies in one paper.
\citet{paper97}, for example, conducted an online study in which they did not find an impact of the participants with a maximizing decision-making strategy on the perceived explanation quality.
In their eye-tracking study, such an impact was found, though.
The differences in results can be caused by the different participants and the change in study types or how the effect was measured (questionnaire compared to eye-tracking recording).

In other instances, the different results are due to different independent variables in the evaluation.
We take this as an indication that other factors must be considered when interpreting the results. 

In this survey, we looked at four potential factors, namely, the \emph{explanation modality}, the \emph{explanation generation method}, the \emph{application domain} in which the explanation was evaluated, and the \emph{type of recommender system}.
Explanations might affect a user depending on their modalities or complexity, as different users might have other preferences.
We could not observe any general patterns that certain explanation types are more likely to impact the results from our corpus.
Possible reasons could be the small number of overall results or the necessity for a more fine-grained categorization of the explanation types.
The impact of a user's agreeableness on the perceived explanation quality, for example, was investigated by the same research team in two separate publications \cite{paper27, paper65} focusing on privacy concerns regarding the information disclosed in the explanation.
Both use text explanations in a group recommendation setting applied to the point-of-interest domain.
The main difference in the explanations that arise from the articles is that the explanation in \citet{paper27} is static and discloses potentially private information, while the explanation in \citet{paper65} is presented in a chat interface that allows the user to interactively decide how much information is revealed to the group.
Given the small data sample that we have for these results, applying a more fine-grained classification of explanation types would further increase the sparsity of the data.

The articles of our corpus often did not disclose information on the underlying recommender system and explanation generation approach.
One possible reason for this could be the focus of the articles. 
Describing the setup of a user study and the characteristics of the participants in sufficient detail requires space, which is usually restricted, especially in conference publications. 
This might have led to less detailed descriptions of the explanation method and underlying recommender systems.
We furthermore identified a set of articles in which the recommendations and corresponding explanations were not generated by an actual recommender system but instead simulated for the user study (see \cite{paper25, paper26, paper27, paper53, paper68, paper97}, for example).
These publications usually focus on investigating how different explanations impact the users and are often published in human-computer-interaction-centered venues such as CHI, UMAP, or IUI.
In future work, it would be interesting to see if these proposed explanations and results can be replicated with existing explanation methods.
In the articles that provided information on the recommendation approach, we found a dominance of content-based and hybrid methods.
This could indicate a trend toward providing information in the explanations that go beyond a user's previous interactions or user-/item-similarity. 
\citet{paper38}, for example, showed that their baseline consisting of a short movie description is often perceived en par, if not better, than human-generated explanations for movie recommendations, indicating that providing additional information on the recommended items can be useful to achieve certain explanation goals.

In terms of application domains, the majority carries a low risk when following recommendations without understanding them.
The number of findings measured in a point-of-interest domain is higher than the number in movie recommendations, but overall, the distribution of domains could be correlated with the domains of the typical datasets used in recommender systems evaluation \cite{chin_datasets_dilemma}. 
We further observe that only a few user characteristics were evaluated in three or more domains, namely gender, need for cognition, neuroticism, domain experience, technical expertise, and visualization literacy.
For other characteristics, we either only have a small number of results or the characteristic was only evaluated in one or two domains.
One reason why the diversity of domains is higher for experience characteristics than for personality could be that capturing the personality of the participants usually goes along with additional questionnaires such as the Big Five, which might increase the risk of survey fatigue. As pointed out in Section \ref{subsubsec:experience}, while there are standardized questionnaires to measure the experience for some domains, it is mostly captured by one item in the questionnaire.

After categorizing the explanations, we still find different results for explanations with the same properties according to this classification, sometimes even from the same article.
\citet{paper87}, for example, compared 11 different explanations and found that the visualization literacy of the participants impacts the quality perception of some explanations while others do not.
Both explanations are visual explanations generated by an intrinsically explainable hybrid recommender system. While an effect was found for their Venn diagram explanation, it was not found for their cluster dendrogram. We leave investigating the detailed differences of such occurrences to determine the cause of these differences for future work.

When examining the papers, we noticed differences in how the effects were defined and measured.
For the effectiveness of explanations, for example, we found papers in which the participant was asked to rate an explanation's or system's effectiveness. In contrast, other papers measured it implicitly (e.g., by evaluating the participants' decision quality).
Further specifying the effect that was measured might show a more precise picture in the cases where mixed results were found.

\subsection{What are the implications of not measuring the impact of user characteristics on explanations?}

Fifteen papers in our corpus did not report any of the user characteristics we looked at in this work, which harms the reproducibility of the results as it is impossible to know whether the recruited participants are similar to the ones in the original study.
Another aspect that makes reproducing and comparing the reported results between the individual papers challenging is how they are described.
Figure \ref{fig:participant-age} illustrates the example of the participant's age and how different the characteristics are reported in our corpus.
This applies to demographics, participants' prior experience, and certain personality traits where the scales and calibration differ between studies.

In our survey, we only found a few studies in which the impact of user characteristics on explanation effects was evaluated.
In this section, we discuss the possible implications of not measuring the impact of user characteristics on explanation effects.

We do not know which user characteristics can influence the effect of explanations on a user.
Therefore, we also do not know what information we need about the users of an application to provide them with an explanation that has the desired effect.
Does it matter what age group the users are?
Do users in their 20s have a different perception of the explanations than users in their 30s? Does the perception change with a bigger age gap?
Despite having the desired effect on most participants, we cannot rule out that the evaluated explanation negatively impacts the user experience of a substantial part of the target users by only assessing the effect of explanations on all participants.

We further noticed that the user characteristics are often evaluated in isolation.
We only found rare occurrences in which multiple user characteristics are grouped.
In fact, only one publication in our corpus disaggregated their results based on two demographic categories.
\citet{paper111} grouped gender with age and evaluated if the explanation evokes a different quality perception of the recommender system between genders in different age groups and education levels.
In all cases, they found an impact on the evaluated explanation effects. Still, it is unclear to what extent forming these subgroups based on multiple characteristics is meaningful and which characteristics should be grouped in the evaluation as we only have this one paper in our corpus.

Knowing which user characteristics to pay attention to could also lead to a more effective and efficient evaluation.
The participants can be targeted more precisely, and the data that needs to be collected about the participants could be narrowed down to the essential aspects.

Ultimately, finding out which user characteristics matter in the evaluation process of explanations requires a bigger focus on reproducibility studies.
In our corpus of 124 papers, only \citet{paper21} reproduced a study.
Comparing the setups of both the original study and the reproduced one might lead to insights on aspects that explain the differences in results.
However, to get to this point, effort needs to be put into increasing the comparability of the studies.

Given our findings, we provide a set of recommendations for the field to be able to move in this direction.

\subsection{Recommendations}
\label{sec:discuss-recs}

Finally, we derive the following recommendations from the results regarding the participant recruiting, reporting of their data, and the evaluation of explanation effects in recommender systems.

\subsubsection{Participant Recruiting}
In terms of the recruitment process, we recommend a \emph{greater awareness of who is being recruited and whether this sample is representative of the target audience of the application domain}.
Furthermore, the \emph{shortcomings of the participant pool in crowdsourcing platforms need to be considered and addressed when opting for this type of recruiting}.
We suggest including a statement in the publication on what efforts were undertaken to ensure a representative participant sample with respect to the application domain.
While the majority of application domains in our corpus have a low risk when a user follows the recommendations without understanding them, they usually target a diverse audience, which should be represented in the evaluation process. Domains such as the medical domain involve a higher risk compared to, e.g., POI recommendations. We, therefore, suggest paying special attention to the representativeness of the participant sample in these domains to make sure that the explanations are beneficial to all users.
We particularly recommend paying attention to the participants' location so as to not exclude users from the Global South.

\subsubsection{Reporting of Participant Data}
The differences in what and how participant data is reported, pointed out in Section \ref{sec:results}, make it impossible to reproduce experiments or draw conclusions about the external validity of the results.
Therefore, we recommend \emph{working toward and applying reporting standards such as the APA style guideline}~\cite{american2022publication}.
They suggest, for example, avoiding reporting the age of participants with open-ended bins and including options for non-binary genders.
A unified way of describing participants in human-subject experiments would not only improve the assessability of the external validity of the experiment results but also the reproducibility of the study.

Regarding \emph{reporting the participants' personality traits, we want to encourage a more fine-grained division of the participants instead of a median split}.
A comparison of, e.g., people with a very high or low NFC score with the average would be interesting to see.
Furthermore, it would improve the reproducibility of the study if the participant distributions were reported for the personality traits instead of the distribution across two groups after a median split.

\subsubsection{Evaluation of Explanations}
We recommend \emph{more research efforts to investigate the impact of user characteristics on the effects of different explanations}.
Knowing how the characteristics of users and explanations are connected is vital to moving toward explanation designs that benefit the entire user base of an application. 
As can be seen in Table \ref{tab:recorded-characteristics}, the majority of articles in our corpus recorded demographic data about their participants, but the results were only disaggregated for a few studies.
Analyzing the impact of demographic characteristics on the measured effects would, therefore, not require additional data collection efforts and could serve as a starting point to learn more about how explanations impact different user groups.
We further want to encourage research teams to include explanations from related work as baselines in their evaluation to learn more about the external validity of the previously reported results.
Table \ref{tab:charactereffects} can serve as a guide to the kinds of questions that require answering and to identify promising baseline explanations.

\subsubsection{Reproducibility}
Several Information Retrieval and Recommender Systems conferences (e.g., RecSys, SIGIR, ECIR, UMAP) have designated reproducibility tracks or include reproducibility efforts in their call for papers.
Despite the challenges that we identified in this work, we want to encourage the community to reproduce existing work and submit to these venues to discuss the difficulties they are facing with the community and to increase awareness of the issues that still need to be solved.
In our analysis of context variables, we identified that many explanations were simulated or not enough information was provided about the explanation generation process.
Trying to generate the evaluated explanations with existing recommender systems and explainable AI methods could make for an interesting starting point. 

\subsubsection{Pre-Registration of User Studies}
Pre-registering a study protocol is common practice in medicine and is increasingly promoted in social sciences to reduce reporting and publication bias \cite{munafo_2017_manifesto}.
We recommend fostering a discussion among the research community on how to establish and adopt the pre-registration practice for human subject evaluation.

\subsubsection{FAIR Data Sharing}
Lastly, we argue that \emph{the community should adopt a privacy-preserving approach to Findable, Accessible, Interoperable, and Reusable (FAIR) data-sharing practices}~\cite{wilkinson2016fair}
to enable both meta-analyses and comparisons across papers.
We acknowledge that given that the data collected is about human subjects, special care should be given to preserve the subjects' privacy, but we believe that privacy-preserving solutions can be found.

\section{Limitations}
\label{sec:limits}

One of the main limitations of this work is that we are bound to what has been reported in the papers of our corpus.
Some papers might have recorded and evaluated user characteristics but not reported them.
It is more likely that statistically significant results are reported than results where the evaluated effect was not found.
This might affect the results for RQ 2.
However, our work highlights what is known in the research community, so we think the reporting bias did not significantly influence our contribution.

Smaller limitations come from the way certain information is reported in the papers.
In publications with multiple user studies, the information on the participants is occasionally summarized and collectively reported for all studies.
In these cases, we only considered the information once when analyzing the participants for RQ 1.
Regarding the location data, it is sometimes unclear if the country of residence or the participants' nationality is meant.
As most papers collected information about the country of residence, we opted to count these cases to this category.

When creating the corpus for a survey, a selection bias might limit the scope of the papers.
However, we mitigated this by querying library databases covering multiple disciplines and not selecting papers based on venue or number of citations. 
Furthermore, we only analyzed the papers published between 2017 and 2022. 
This timespan showed the largest increase in explainability research~\cite{vultureanualbisi_survey_2022}, demonstrating that the sample of papers we analyzed covers a significant time period. 

Regarding the validity of the results of RQ2, we only considered the impact of user characteristics on explanation effects and the application domain, explanation modality, explainability type, and recommender type as context variables.
However, as indicated in Section~\ref{sec:discuss-charcteristic-impact}, other factors might also play a role, such as the evaluation approach and methodology.
Another factor limiting these results is how we categorized the evaluated explanation effects.
Our goal was to provide an overview of which user characteristics have been evaluated on which effects.
Therefore, we kept the effect groups on a broad level.
However, the publications evaluated different aspects of user experience or other notions of trust.
We plan to extend our analyses of the results in this corpus in future work to include the aforementioned context variables and a more fine-grained categorization of the explanation effects.

Lastly, the publication period of our corpus selection includes the years in which the Covid-19 pandemic made in-person user studies impossible in most parts of the world.
This could have led to an over-proportional representation of online- and crowdsourced studies.
However, these study types were already popular before the start of the pandemic, which is why we do not believe that this had an impact on our results.

\section{Conclusions and Future Work}

In this literature survey, we examined recent publications in which recommender systems explanations were evaluated on users and investigated the question of who these explanations actually serve.

Analyzing the surveyed study participants and answering the first research question, it was brought to light that \emph{most studies do not reflect the average population and exclude larger user groups} such as people from the global south, users without academic degrees, or older people.
Overall, the information on what characteristics are being recorded and how they are reported can vary greatly between publications.
In terms of the second research question, we cannot definitely say if these characteristics impact the effect that explanations have on the users due to the small number of studies in which this was evaluated. Our analysis only shows possible indications that should be further investigated by future work in order to find an answer to this research question.

We close our discussion with recommendations for participant recruiting, reporting of participant data, and evaluation of explanation effects. 
The \emph{recruiting process should be more mindful of reflecting the targeted user group regarding demographics and characteristics}.
A description of the targeted user groups and a statement of how the recruited participants reflect them should be explicitly mentioned.
\emph{The way participant data is reported in the papers should follow standards to ensure comparability across papers and improve their reproducibility}.
Ideally, the data would be shared in a privacy-preserving, FAIR manner to enable subsequent analyses and data re-use.
In order to know which user characteristics matter, we recommend putting more effort into investigating the impact of user characteristics on explanation effects.

In future work, we plan to extend this analysis to address the limitations and gain further insights into the effects of explanations on users and aspects that can impact them. 
We hope this will pave the way to a better understanding of the interaction between user characteristics and explanations in recommender systems --- a topic in dire need of elucidation.

\begin{acks}
This work was partly supported by the Swiss National Science Foundation through project \href{https://data.snf.ch/grants/grant/202125}{``MediaGraph''} (contract no.\ 202125).
\end{acks}

\bibliographystyle{ACM-Reference-Format}
\bibliography{bib}

\end{document}